\documentclass[11pt,twoside,a4paper]{article}
\usepackage{moreverb}
\usepackage{amsmath}
\usepackage{bm}
\usepackage{algorithm,algorithmic}
\usepackage{subfigure}
\usepackage{cuted}
\usepackage{graphicx}

\newcommand\BibTeX{{\rmfamily B\kern-.05em \textsc{i\kern-.025em b}\kern-.08em
T\kern-.1667em\lower.7ex\hbox{E}\kern-.125emX}}
\newcommand{\dt}{h}
\DeclareMathOperator{\tr}{tr}
\newcommand{\textsub}[2]{{#1}_{\text{#2}}}
\newcommand{\textsup}[2]{{#1}^{\text{#2}}}
\makeatother
\makeatletter
\renewcommand*\env@matrix[1][*\c@MaxMatrixCols c]{%
\hskip -\arraycolsep
\let\@ifnextchar\new@ifnextchar
\array{#1}}
\makeatother

\begin{document}


\title{Meshfree elastoplastic solid for nonsmooth multidomain dynamics}

\author{J.~Nordberg and M.~Servin}


\maketitle

\begin{abstract}
A method for simulation of elastoplastic solids in multibody systems with 
nonsmooth and multidomain dynamics is developed. 
The solid is discretised into pseudo-particles using the meshfree moving least squares method.
The particles carry strain and stress tensor variables that are mapped to
deformation constraints and constraint forces.  The discretised solid model thus fit a 
unified framework for nonsmooth multidomain dynamics for realtime simulations
including strong coupling of rigid multibodies with complex kinematic constraints such as articulation joints, unilateral contacts with dry friction, 
drivelines and hydraulics.  The nonsmooth formulation allow for impulses, due to impacts for instance, to 
propagate instantly between the rigid multibody and the solid.
Plasticity is introduced through an associative perfectly plastic modified Drucker-Prager model. 
The elastic and plastic dynamics is verified for simple test systems and the 
capability of simulating tracked terrain vehicles driving on a deformable terrain 
is demonstrated.
\end{abstract}

\section{Introduction}
\label{sec:introduction}
\vspace{-2pt}
We address the modeling and simulation of elastoplastic solids in multidomain environments
including also mechatronic multibody systems with nonsmooth dynamics, such as vehicles,
robots and processing machinery.  Fast multidomain simulation is useful for concept design 
exploration, development of control algorithms and for
interactive realtime simulators, e.g., for operator training, human-machine 
interaction studies and hardware-in-the-loop testing.

Realizing such simulators require integrating many subsystems of different types and 
complexity into a single multidomain dynamics model. If the subsystems are loosely coupled, 
the full system simulation can be realised by means of co-simulation \cite{kubler:2000:tms},
but this is not the general case.  For the sake of computational performance, coarse 
grain models are often used, with rigid and flexible bodies coupled by kinematic 
constraints for modeling of joints and differential algebraic equation (\textsc{dae}) models
for electronics, hydraulics and powertrain dynamics \cite{Burgermeister:2011:sva}. At coarse timescales the dynamics 
need to be treated as nonsmooth, allowing discontinuous velocities of rigid bodies undergoing 
impacts or frictional stick-slip transitions and instantaneous propagation of impulses 
through the system. 
This requires strong coupling through consistent mathematical formulation and coherent numerical treatment of 
the full system in order to achieve good stability and computational efficiency.
Current techniques for co-simulation does not support that.

The theory and numerical methods for nonsmooth multidomain mechanics
is covered in the reference \cite{acary:2008:nmn}. The framework in 
 \cite{lacoursiere07,lacoursiere:2007:rvs} is employed for constraint regularisation and
stabilisation based on a discrete variational approach
with constraints introduced as the stiff limit of energy and dissipation potentials. 

\begin{figure}[h!]
\centering  \includegraphics[width=0.45\textwidth]{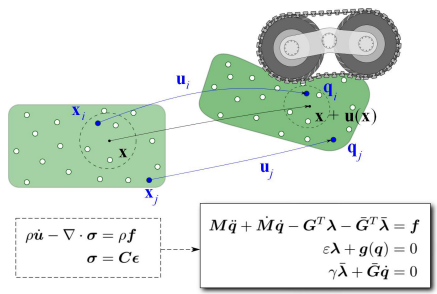}
\caption{Illustration of the idea of using meshless discretisation and
deformation constraints to obtain a unified multibody system model
for both deformable solids and mechatronic systems.}
\label{fig:illustration}
\end{figure}

The developed method is applicable to many different systems but one is primarily explored,
namely the interaction between ground vehicles and deformable terrain \cite{Taheri:2015:tst}.  Current solutions 
enable simulation of complex vehicle models with realtime performance but not including dynamic terrain models
firmly based on solid mechanics in three dimensions.  Usually, empirical terrain models of Bekker-Wong type
are used \cite{Azimi:2015:mdf,Madsen:2013:cbd}.  For more accurate offline simulations with fine temporal 
resolution, on the other hand, many solutions exist for elastoplastic solids coupled with tire models \cite{shoop:2001:fem, xia:2013:fem} but scarcely with more complex multibody systems and not for realtime or faster simulation.

The main contribution in this paper, illustrated in figure \ref{fig:illustration},
is a formulation and numerical method for simulation of 
elastoplastic solids as a nonsmooth multidomain multibody 
system on descriptor form \cite{kubler:2000:tms}.  
This enables numerical integration with large timesteps that potentially exceed current solutions and provide realtime performance or better.

The regularisation and damping terms regulate the elasticity and viscous damping of deformations 
according to linear elasticity theory. An associative perfectly plastic  
Drucker-Prager model is employed using an elastic predictor-plastic corrector strategy to detect yielding and compute the plastic flow. In its basic form this model does not yield in hydrostatic compression
in contrast to many real materials. Many soils yield under hydrostatic compression
by failure in the microscopic structures whereby air and fluid is released.
Therefore the Drucker-Prager model is extended to a capped version \cite{Dolarevic:2007}. 

The dynamics of other subsystems and their potentially strong coupling 
is treated within the same multibody dynamics framework using a variational time 
integrator.  Each timestep involve solving a block-sparse  mixed linear complementarity problem \textsc{mlcp}, and can be efficiently integrated with tailored solvers.

A meshfree method \cite{liu:2010:imm,Nguyena:2008:mmr} is chosen in order to handle large deformations without need for
remeshing \cite{Ullah:2013:fde} and, for future development, support fracturing and transitions to 
viscous flow or to granular media represented by contacting discrete elements.
The displacement gradient and strain tensor is approximated by the method of moving least
squares (\textsc{mls}) \cite{Belytschko:1994:efg}.

\subsection{Notations}
\label{sec:notations}
Matrices and vectors are represented in bold face in capital and lower case, 
$\bm{A}$ and $\bm{x}$, respectively. Latin superscripts indicate the index of a specific 
body $i,j, k = 1,\hdots, N$, where $N$ is the total number of bodies in the system.
Greek subscripts indicate a specific scalar component of a vector or matrix
(may be multidimensional). The Einstein summation convention is used
where repeated indices imply summation over them, e.g., representing
matrix-vector multiplication as $\bm{A}\bm{x} = A_{\alpha \beta} x_\beta$. 
Subscripts $x, y$ or $z$ indicate a specific component of a vector or matrix
assuming a Cartesian coordinate system. As an example of these notations  the position vector of a body $i$ is denoted $\bm{q}^{i}$.  
The relative position of particle $j$ to $i$ is 
$\bm{q}^{ij} = \bm{q}^{j} - \bm{q}^{i}$.  
The $y$ component of this vector is $q^{ij}_{y} = q^j_y - q^i_y$.  
For rigid bodies, the position vector include also the rotation of the body.  
The system position vector is 
$\bm{q} = [\bm{q}^1, \bm{q}^2,\hdots, \bm{q}^i, \hdots, \bm{q}^N]$.
The gradient is denoted $\nabla_{\bm{x}} = \frac{\partial}{\partial \bm{x}}$
and $\nabla_{\alpha} = \frac{\partial}{\partial x_{\alpha}}$.
The dot notation is used for the time derivative $\dot{} \equiv \tfrac{\text{d}}{\text{d}t}$.

\vspace{-6pt}
\section{Discrete multidomain dynamics}
\label{sec:discrete_multidomain}
\vspace{-2pt}

\subsection{Lagrangian formulation}
\label{sec:lagrangian}
The Lagrangian of a constrained mechanical system is
\begin{align}
  \label{eq:lagrangian}
  \mathcal{L}(\bm{q},\dot{\bm{q}},\bm{\lambda},\bar{\bm{\lambda}}) & = 
  T(\bm{q},\dot{\bm{q}}) - U(\bm{q}) - \mathcal{R}(\bm{q},\dot{\bm{q}}) \nonumber\\
 & + \bm{\lambda}^T \bm{g}(\bm{q},t) + \bar{\bm{\lambda}}^T \bar{\bm{g}}(\bm{q},\dot{\bm{q}},t)
\end{align}
where $T(\bm{q},\dot{\bm{q}}) = \tfrac{1}{2}\dot{\bm{q}}^T \bm{M}\dot{\bm{q}}$ is the 
kinetic energy, 
$U(\bm{q})$ is the potential energy and $\mathcal{R}(\bm{q},\dot{\bm{q}})$ 
is the Rayleigh dissipation function.  The system
state vector $(\bm{q},\dot{\bm{q}})$ is constrained by scleronomic holonomic constraints
$0 = \bm{g}(\bm{q})$, with Jacobian $\bm{G} = \partial \bm{g} / \partial \bm{q}$,
and nonholonomic Pfaffian constraints 
$0 = \bar{\bm{g}}(\bm{q},\dot{\bm{q}},t) \equiv \bar{\bm{G}} \dot{\bm{q}} - \bm{w}(t)$
with corresponding Lagrange multipliers $\bm{\lambda}$ and $\bar{\bm{\lambda}}$.

To ensure numerical stability it is common to regularise the
constraints. This may be done by treating them as the limit
of strong potentials and dissipation functions
$U_{\varepsilon}(\bm{q}) = \tfrac{1}{2\varepsilon}\bm{g}^T\bm{g}$
and $\mathcal{R}_\gamma(\bm{q},\dot{\bm{q}}) = \tfrac{1}{2\gamma}\bar{\bm{g}}^T\bar{\bm{g}}$
with $\varepsilon,\gamma \to 0$ \cite{bornemann:1997:hhs}.
In fact, any stiff force can be transformed into
a regularised constraint through a Legendre transform
\cite{lacoursiere:2007:rvs} 
\begin{align}
  \label{eq:potentials}
  U_{\varepsilon}(\bm{q}) & = \bm{\lambda}^T\bm{g} 
  - \tfrac{\varepsilon}{2}\bm{\lambda}^T\bm{\lambda}\\
  \mathcal{R}_\gamma(\bm{q},\dot{\bm{q}}) & = 
  \bar{\bm{\lambda}}^T\bar{\bm{g}} - 
  \tfrac{\gamma}{2}\bar{\bm{\lambda}}^T\bar{\bm{\lambda}} 
\end{align}
This transform the Euler-Lagrange equations of motion into
\begin{align}
  \label{eq:euler_lagrange}
  \bm{M}\ddot{\bm{q}} + \dot{\bm{M}}\dot{\bm{q}}
  - \bm{G}(\bm{q})^T \bm{\lambda} - \bar{\bm{G}}(\bm{q})^T \bar{\bm{\lambda}}& = \bm{f}\\
  \varepsilon \bm{\lambda} + \bm{g}(\bm{q}) & = 0\label{eq:reg_holonomic}\\
  \gamma \bar{\bm{\lambda}} + \bar{\bm{G}}(\bm{q})\dot{\bm{q}} & = \bm{w}(t)
  \label{eq:reg_nonholonomic}
\end{align}
where $\bm{f} \equiv -\nabla_{\bm{q}}U(\bm{q}) -\nabla_{\dot{\bm{q}}}\mathcal{R}(\bm{q},
\dot{\bm{q}})$ are the explicit forces from weak smooth potentials.
This constitute a system of \textsc{dae} of index 1, which are easier to integrate 
numerically than the corresponding higher index \textsc{dae} in the absence of
 regularisation and stabilisation. $\varepsilon = \gamma = 0$.
A term for dissipation of motion orthogonal to the holonomic 
constraint surface can be added to improve the convergence.
Also this dissipation can be physically based
by introducing it as a Legendre transform of a
Rayleigh dissipation function
$\mathcal{R}_\tau(\bm{q},\dot{\bm{q}}) 
= \tfrac{\tau}{2\varepsilon}\dot{\bm{g}}^T\dot{\bm{g}}
\to \tau \dot{\bm{\lambda}}^T\dot{\bm{g}} 
- \tfrac{\tau\varepsilon}{2}\dot{\bm{\lambda}}^T\dot{\bm{\lambda}}$,
with damping parameter $\tau$.  This modifies Eq.~\eqref{eq:reg_holonomic} to 
\begin{equation}
\varepsilon \bm{\lambda} + \varepsilon\tau \dot{\bm{\lambda}} 
  + \bm{g}(\bm{q})  +  \tau \bm{G}\dot{\bm{q}} = 0 
\end{equation}

The compliance and damping factors $\varepsilon$, $\tau$ and $\gamma$
are not restricted to being scalar or diagonal.  In what follows
these are assumed to be matrices.

\subsection{Time discretisation}
\label{sec:time_discretisation}
Variational integration \cite{Marsden:2001:DMVb} provides a systematic approach
to derive time integration schemes with good properties, e.g., momentum preservation and 
symplecticity.
Rather than discretising the Euler-Lagrange equations of motion directly,
the Lagrangian and principle of least action is defined in time-discrete form.
Employing semi-implicit Euler discretisation and linearising the 
constraint as
$\bm{g}(\bm{q} + \Delta \bm{q}) = \bm{g}(\bm{q}) + \bm{G} \Delta \bm{q}$
leads to the following scheme
\begin{align}
  \bm{q}_{n+1} & = \bm{q}_{n} + h \dot{\bm{q}}_{n+1}
\end{align}
\begin{align}
  \label{eq:saddle_point}
  \begin{bmatrix} 
  \bm{M} & -\bm{G}^T & -\bar{\bm{G}}^T\\ 
  \bm{G} & \bm{\Sigma} & 0\\ 
  \bar{\bm{G}} & 0 & \bar{\bm{\Sigma}}
  \end{bmatrix}
  \begin{bmatrix} \dot{\bm{q}}_{n+1} \\ \bm{\lambda} \\ \bar{\bm{\lambda}} \end{bmatrix} 
  = \begin{bmatrix} 
  \bm{p}_n \\ 
  \bm{v}_n \\ 
  \bm{\omega}_n 
  \end{bmatrix}
\end{align}
where $\bm{p}_n =  M\dot{\bm{q}}_n + \dt \bm{f}_n$, $\bm{v}_n =  -\frac{4}{\dt}\Upsilon g + \Upsilon G\dot{\bm{q}}_n$
and regularisation and stabilisation matrices  
$\bm{\Sigma} = \text{diag}(4\varepsilon/\dt^2(1+4\tau/\dt))$, $\bar{\bm{\Sigma}} = \text{diag}(\gamma/\dt)$ and $\bm{\Upsilon} = \text{diag}(1 + 4 \tau/\dt)^{-1})$.
Eq.~\eqref{eq:saddle_point} is a linear system of $N = \dim(\dot{\bm{q}}) + \dim(\bm{g}) + 
\dim(\bar{\bm{g}})$ equations.
The matrix on the left hand side is block-sparse, positive definite and non-symmetric.
The regularisation appearing as the diagonal perturbation matrices $\bm{\Sigma}$ and $
\bar{\bm{\Sigma}}$ are needed for handling otherwise ill-posed or ill-conditioned problems, e.g., 
systems with constraint degeneracy and large mass ratios.
The stabilisation terms $-\frac{4}{\dt}\Upsilon g + \Upsilon G\dot{\bm{q}}_n$ on the right 
hand side counteract constraint violations, e.g., sudden and large contact penetrations at 
impacts or small numerical constraint drift.
The presented stepping scheme, referred to as \textsc{spook}, has been proved to be linearly
stable \cite{lacoursiere07} and numerical simulations suggest a large domain of nonlinear 
stability.

The combined effect of the regularisation and stabilisation terms is to
bring elastic and viscous properties to motion orthogonal to the constraint
surface $\bm{g}(\bm{q}) = 0$, e.g., for modeling elasticity in mechanical joints.
The parameters $\varepsilon$ and $\gamma$ need not be chosen arbitrarily,
as for the conventional approaches to constraint regularisation and
stabilisation, but can be based on physics models using parameters that can be 
derived from first principles, found in literature or be identified by 
experiments. This becomes straightforward when the regularisation is introduced 
by potential energy as quadratic functions in $\bm{g}$, i.e., 
$U_{\varepsilon}(\bm{q}) = \tfrac{1}{2}\bm{g}^T\bm{\varepsilon}^{-1}\bm{g}$.
This has been exploited in previous work to constraint based 
modeling of lumped element beams \cite{servin:2007:rbc}, wires \cite{servin:2010:hmw},
meshfree fluids \cite{bodin:2010:cf} and granular material \cite{servin:2014:esn}.

\subsection{nonsmooth dynamics}
\label{sec:nonsmooth}
In discrete time, some of the dynamics is best treated as 
\emph{nonsmooth} \cite{acary:2008:nmn},
meaning that the velocity may change discontinuously to fulfil inequality
constraints and complementarity conditions.
Introducing such conditions is an efficient way of modeling 
contacts, dry friction, joint limits, electric circuit switching,
motor and actuator dynamics and control systems for discrete time with large 
step size.  The alternative would be to use a fine enough 
temporal resolution where the dynamics appear smooth 
and may be modeled by \textsc{dae}s or ordinary differential 
equations alone.  In general, this is an intractable approach for full system 
simulations and interactive applications.
The nonsmooth dynamics can be formally treated as \emph{differential
variational inequalities} \cite{Pang:2008:DVI}. The discrete equations of motion \eqref{eq:saddle_point} take the form of a \emph{mixed linear complementarity problem}, \textsc{mlcp}, when nonsmooth dynamics is included
\begin{align}
  \label{eq:mlcp}
  \bm{H}\bm{z}+\bm{r} & = \bm{w}_+ - \bm{w}_-\\
  0 \leq \bm{w}_+ & \perp \bm{z}-\bm{l} \geq 0\nonumber\\
  0 \leq \bm{w}_- & \perp \bm{u}-\bm{z} \geq 0\nonumber
\end{align}
where $\bm{H}\bm{z}$ corresponds to the left hand side of equation \eqref{eq:saddle_point}, 
$\bm{r}$ is the negated right hand side and $\bm{w}_\pm$ are slack variables. 
The terms $\bm{u}$ and $\bm{l}$ correspond to the upper and lower limits on the 
solution vector $\bm{z}$, respectively.  The original linear system of
equations is recovered by assigning $\pm \infty$ to the upper and lower limits,
 i.e., no limit.

Rigid body contacts are handled by the Signorini-Coulomb law for unilateral
dry frictional contacts.  This states that if the non-penetration 
constraint, $g\geq 0$, is violated then the normal contact velocity
and constraint force are complementary in ensuring separation,
$0 \leq \textsub{v}{n} \perp \textsub{\lambda}{n} \geq 0$,
and the tangential friction force, acting to maintain zero slip,
is bound to be on or inside the Coulomb friction cone, 
$|\textsub{\bm{\lambda}}{t}| \leq \mu \textsub{\lambda}{n}$. 
The latter may be linearised by approximating the cone with a polyhedral or a box.
Impacts are treated post facto in a separate \emph{impact stage}
succeeding the update of velocities and positions.  At the impact stage
an impulse transfer is applied enabling discontinuous velocity changes, i.e., from $\dot{\bm{q}}^{−}$ to $\dot{\bm{q}}^+$. 
Newtons' impact law,
 $\bm{n}^T\dot{\bm{q}}^{+} = - e \bm{n}^T\dot{\bm{q}}^-$,
is used with restitution coefficient $e$ in conjunction with preserving 
all other constraints, 
$\bm{G}\dot{\bm{q}} = 0$.  This amounts to solving the same \textsc{mlcp} \eqref{eq:mlcp} 
but with $\dt \bm{f}_n = \bm{0}$ and 
$-\frac{4}{\dt}\Upsilon g + \Upsilon G\dot{\bm{q}}_n = \bm{0}$ 
in the right hand side of equation \eqref{eq:saddle_point}.

A fixed timestep approach is preferred when aiming for fast 
full systems simulations with many nonsmooth events with
inequality or complementarity conditions, e.g.,
involving thousands or millions of dynamic contacts,
switching in electric, hydraulic or control systems,
or elastic material undergoing fracture or plastic failure.
Using variable timestep and exact event location become
computationally intractable and may fail by occurrence of
Zeno points.  

\subsection{Heterogeneous multidomain dynamics}
In a multidomain simulation with strongly 
coupled subsystems the nonsmooth dynamics propagate instantly 
throughout the full system. The dynamics in the
elastoplastic solid may be directly and strongly coupled
with the internal dynamics of a mechatronical system.
The nonsmooth multidomain dynamics approach with physics based
constraint regularisation and stabilisation presented here,
provides a general framework for building and efficiently
solving the dynamics on \textsc{mlcp} form.  Each subsystem take
the same generic form of saddle-point matrix
as the full system in Eq.~\eqref{eq:saddle_point}.
Considering a system with two subsystems $A$
and $B$, the full coupled system takes the following form
\begin{align}
	\bm{H}\bm{z} & = 
  \begin{bmatrix}[cc|c] 
  \bm{H}_{A} & 0 & -\bm{G}^T_{{AB}_A}\\ 
  0 & \bm{H}_{B} & -\bm{G}^T_{{AB}_B}\\
  \hline
  \bm{G}_{{AB}_A} & \bm{G}_{{AB}_B} & \bm{\Sigma}_{AB}
  \end{bmatrix}
  \begin{bmatrix} \bm{z}_{A} \\ \bm{z}_B \\ \bm{\lambda}_{AB} \end{bmatrix}
  \label{eq:self_similar}
\end{align}
where $\bm{\lambda}_{AB},[\bm{G}_{{AB}_A},\bm{G}_{{AB}_B}],\bm{\Sigma}_{AB}$ are the multiplier, Jacobian and regularisation of the subsystem coupling. 
\section{Constraint based meshfree elastoplastic solid}
\label{sec:constraint_elastoplastic}

\subsection{Elastoplastic solid}
\label{sec:elastoplastic}
The solid is assumed to sustain geometric large deformations.
The St.~Venant-Kirchoff elasticity model is used in combination with the 
Drucker-Prager plasticity model \cite{Neto:CMP:2008}.
The material strain is expressed by the Green-Lagrange strain tensor
\begin{align}
  \label{eq:greenstvenant}
  \epsilon_{\alpha\beta}(\bm{x}) & = 
  \frac{1}{2} \left( \nabla_{\alpha} u_\beta + \nabla_{\beta} u_\alpha + 
  \nabla_{\alpha} u_\gamma\nabla_{\beta} u_\gamma \right)
\end{align}
where $\bm{u}(\bm{x})$ is the displacement field mapping a reference coordinate $\bm{x}$
to displaced position $\tilde{\bm{x}}(\bm{x}) = \bm{x} + \bm{u}(\bm{x})$.
Observe that the Green-Lagrange strain tensor transforms
under large rotations and no co-rotation procedure is needed.
It is convenient to use the Voigt notation for representing the strain tensor
on vector format, $\bm{\epsilon} = \left[ \epsilon_{xx}, \epsilon_{yy}, \epsilon_{zz}, 
2\epsilon_{yz}, 2\epsilon_{xz}, 2\epsilon_{xy} \right]^T$, and the stress tensor
$\bm{\sigma} = \left[ \sigma_{xx}, \sigma_{yy}, \sigma_{zz}, 
\sigma_{yz}, \sigma_{xz}, \sigma_{xy} \right]^T$ such that the linear
constitutive law reads 
$\bm{\sigma} = \bm{C} \bm{\epsilon}$, with stiffness matrix 
\begin{align}
  \label{eq:stifnesstensorreduced}
  \bm{C} = \begin{bmatrix}\lambda + 2\mu & \lambda & \lambda & 0 & 0 & 0\\
  \lambda & \lambda + 2\mu & \lambda & 0 & 0 & 0\\
  \lambda & \lambda & \lambda + 2\mu & 0 & 0 & 0\\
  0 & 0 & 0 & \mu & 0 & 0\\
  0 & 0 & 0 & 0 & \mu & 0\\
  0 & 0 & 0 & 0 & 0 & \mu
  \end{bmatrix}
\end{align}
where $\lambda$ and $\mu$ are the first and second Lam\'{e} parameters which are 
related to the Young's modulus and Poisson's ratio, $E$ and $\nu$, of a material as
$ \lambda = E\nu/\left(1+\nu\right)\left(1-2\nu\right)$ and
$\mu = E/2\left( 1+\nu \right)$. 
The corresponding strain energy density is
\begin{align}
  \label{eq:strainenergy}
  U(\bm{x}) = \frac{1}{2} \bm{\epsilon}^T \bm{C} \bm{\epsilon} = \lambda (\tr
  \bm{\epsilon})^2 + 2\mu \tr( \bm{\epsilon}^2)
\end{align}
Plastic deformation occur when the stress of a material reaches its yield strength, 
$\Phi(\bm{\sigma}) = 0$, in which case the material undergo plastic flow if the 
load is increased further.
The choice of yield function, $\Phi(\bm{\sigma})$, 
is based on the type of material being modeled.
To represent the permanent plastic deformation, 
the strain tensor is decomposed into an elastic and a plastic component, 
$\bm{\epsilon} = \bm{\epsilon}^{\text{e}} + \bm{\epsilon}^{\text{p}}$. In
ideal plasticity the plastic deformation occurs instantly according to a flow rule
$\text{d}\bm{\epsilon}^{\text{p}} = \text{d}\bm{\lambda}^{\text{p}} 
\frac{\partial \Psi}{\partial \bm{\sigma}}$,
where $\Psi(\bm{\sigma})$ is the plastic potential and $\bm{\lambda}^{\text{p}}$ is 
the plastic multiplier. If $\Psi(\bm{\sigma}) = \Phi(\bm{\sigma})$
the model is said to be associative and otherwise non-associative.
The plastic flow last as long as it is positive, 
$\text{d}\bm{\lambda}^{\text{p}}>0$, incrementally reducing the stress 
$\text{d}\bm{\sigma}^{\text{p}}$, until it reaches the elastic regime, 
$\Phi(\sigma) < 0$. This constitute a nonlinear complementarity problem known as 
Karush-Kuhn-Tucker conditions
\begin{align}
  \Phi \leq 0,\;\; \text{d}\bm{\lambda}^{\text{p}} \geq 0,\;\; \Phi\ \text{d}
  \bm{\lambda}^{\text{p}} = 0
\end{align}
The plastic multiplier is computed from the constitutive law, which in
the plastic flow phase is $\text{d}\bm{\sigma} =\textsup{\bm{C}}{p} \text{d}\bm{\epsilon}$,
where the elastoplastic tangent stiffness matrix is
\begin{equation}
  \label{eq:constitutive_matrix}
 	\textsup{\bm{C}}{p} = \bm{C} - \frac{\bm{C}\frac{\partial\Psi}{\partial\bm{\sigma}}
 	(\frac{\partial\Phi}{\partial\bm{\sigma}})^T\bm{C}}
 	{(\frac{\partial\Phi}{\partial\bm{\sigma}})^T
 	\bm{C} \frac{\partial\Psi}{\partial\bm{\sigma}}} 
\end{equation}
Predictor-corrector algorithms are conventionally used to integrate the plastic
flow.  In the absence of plastic 
hardening or softening, the plastic deformation, plastic multiplier and total stress can 
be computed  easily using the radial return algorithm \cite{Neto:CMP:2008} summarised in 
Algorithm \ref{alg:radial_return}.  

The assumed plasticity model is a capped Drucker-Prager model, 
following Dolarevic \cite{Dolarevic:2007}, with a compaction variable $\kappa$.
The yield function $\Phi\left( \bm{\sigma},\kappa \right)$ is a piecewise function 
consisting of the Drucker-Prager, tension and compression cap functions 
according to \eqref{eq:caps}
\begin{align}
\label{eq:caps}
  \Phi\left( \bm{\sigma},\kappa \right) = 
  \begin{cases}
    \Phi_{\textsc{\tiny T}}\left( I_1,J_2 \right) &  I_1 \geq I_1^{\textsc{\tiny T}} \\
    \Phi_{\textsc{\tiny DP}}\left( I_1,J_2 \right) & I_1^{\textsc{\tiny T}} > I_1 > I_1^{\textsc{\tiny C}}\left( \kappa \right)\\
    \Phi_{\textsc{\tiny C}}\left( I_1 , J_2,\kappa\right) &  I_1^{\textsc{\tiny C}}\left( \kappa \right) \geq I_1 
  \end{cases}
\end{align}
where $I_1 = \tr(\bm{\sigma})$ is the first invariant of the stress tensor and
$J_2 = 0.5\tr(\bar{\bm{\sigma}}^2)$ is the second
invariant of the deviatoric stress tensor
$\bar{\bm{\sigma}} = \bm{\sigma} - \tfrac{1}{3} I_1 \bm{1}$. The expressions for the tension cap 
$\Phi_{\textsc{\tiny T}}\left( I_1,J_2 \right)$ and compression cap 
$\Phi_{\textsc{\tiny C}}\left( I_1 , J_2,\kappa\right)$ are given in Appendix A and the Drucker-Prager surface is defined as
\begin{equation}
\label{eq:DP}
  \Phi_{\textsc{\tiny DP}}\left( I_1,J_2 \right)  = \sqrt{J_2 } + \frac{\eta}{3} I_1 - \xi c 
\end{equation}
where $\phi$ is the internal 
friction angle and $c$ the cohesion parameter such that $\eta = 6\sin \phi/\sqrt{3}\left( 3-\sin \phi \right)$ and
$\xi = 6\cos \phi/\sqrt{3}\left( 3-\sin \phi \right)$.
 
\begin{figure}[h!]
\centering
  \includegraphics[width=0.45\textwidth]{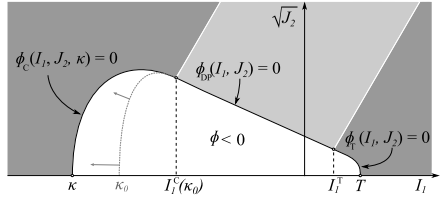}
\caption{The capped Drucker-Prager yield surface.  The non-shaded region
indicate the elastic domain, which grows under plastic compression.}
\label{fig:gapfunction}
\end{figure}

The capped yield surface is illustrated in Figure \ref{fig:gapfunction}.
The tension cap is fix and is used to regularise the plastic flow behaviour 
in the corner region.  The compression surface cap, on the other hand, is dynamic
and the maximum hydrostatic pressure, $\kappa$, is used as main variable.
The conventional Drucker-Prager model is a common model for the plastic deformation
dynamics of soils, e.g. wet or dry sand.   These materials are weak under tensile stress ($I_1 > 0$) and
become stronger under compression ($I_1 <0$) where it may support large
shear stresses ($\sqrt{J_2}$).  The capped Drucker-Prager model is a generalisation that
include also plastic compaction that may occur in many soils.  The compaction mechanism
may be failure of the individual grains whereby air or water is displaced from the soil.  The
compaction saturates at a maximum level, where all voids are vanished.  The compaction dynamics
is modeled by a variable cap on compressive side of the Drucker-Prager yield surface, intersecting
the hydrostatic axis at $-\kappa$.  The compaction hardening law is chosen
\begin{equation}
  \label{eq:kappa}
  \kappa = \kappa_0 + \frac{1}{D}\ln\left( 1+\frac{\tr(\textsup{\bm{\epsilon}}{p})}{W} 
  \right)
\end{equation}
where $W$ is the maximum volume compaction, $D$ is the hardening 
rate and $\kappa_0$ the initial cap position where compressive failure first
occur.  Observe that when the plastic volume compaction $-\tr(\textsup{\bm{\epsilon}}{p})$
approaches $W$, the cap variable $\kappa$ goes to infinity, the material
does not compact further plastically and behave like the standard Drucker-Prager model.
The elastoplastic material parameters are summarised in Table
\ref{tab:elastoplastic}.  Expressions for the detailed shape of the compression cap and 
the derivatives of the yield functions are found in Appendix A.  
\begin{table}
  \centering
  \caption{Elastoplastic model parameters}
  \label{tab:elastoplastic}
  \begin{tabular}{lr}
    \hline
    $E, \nu$ & Young's modulus and Poisson's ratio\\
    $\phi, c$ & friction angle and cohesion\\
    $W, D$ & max compaction and hardening rate\\
    $T, \kappa$ & Cap $I_1$-intersections\\
    \hline
  \end{tabular}
\end{table}

\subsection{Meshfree method}
\label{sec:meshfree}
The continuous solid of mass $m$ and reference volume $V$ 
is discretised into $\textsub{N}{p}$ particles. The particles have mass $m^i = m /\textsub{N}{p}$, 
volume $V^i = V /\textsub{N}{p}$, position $\bm{q}^i$ and velocity $\dot{\bm{q}}^i$.  
The particle displacement is $\bm{u}^i = \bm{q}^i - \bm{x}^i$ with reference position $\bm{x}^i$ as illustrated in figure \ref{fig:illustration}.
A continuous and differentiable displacement field is approximated using the \textsc{mls} method
\cite{Belytschko:1994:efg,wasfy:2003:csf}. The differentiated displacement field defines 
the strain tensor field in any point as described in 
equation \eqref{eq:greenstvenant}.  The particles can thus be understood as pseudo-particles 
having both particle and field properties.  

The \textsc{mls} approximation of the displacement field is
\begin{equation}
\label{eq:displacement_field}
	u_\alpha (\bm{x}) = \sum_j^{\textsub{N}{p}} \Psi_j(\bm{x}) u^j_\alpha
\end{equation}
where the shape function and moment matrix use a quadratic basis
\begin{align}
	\Psi_j(\bm{x}) &= p_\gamma (\bm{x}) A_{\gamma\tau}^{-1} (\bm{x}) p_\tau (\bm{x}^j) W\left(\bm{x} - \bm{x}^j\right)\nonumber\\
	A_{\gamma\tau} (\bm{x}) & = \sum_j^{\textsub{N}{p}} W\left(\bm{x} - \bm{x}^j\right) p_\gamma^j p_\tau^j \\
	\bm{p} (\bm{x}) & = \left[ 1,\  x,\  y,\
  z,\ yz,\ xz,\ xy,\ x^2,\ y^2,\ z^2 \right]^T´\nonumber
\end{align}
and weight function $ W(\bm{x}) = \frac{315}{64\pi l^9} (l^2 - \bm{x}^T\bm{x} )^3$ if 
$|\bm{x}| $is smaller than the influence radius $l$, and otherwise zero.  The base function notation $\bm{p}^j = \bm{p} (\bm{x}^j)$ is used to simplify the expressions.
The gradient of the interpolated displacement field is 
\begin{align}
\label{eq:gradient}
\nabla_\beta u_\alpha (\bm{x}) &= \sum_j^{\textsub{N}{p}} \nabla_\beta \Psi_j(\bm{x}) u^j_\alpha
\end{align}
\begin{align}
\label{eq:gradientPsi}
\nabla_\beta \Psi_j = & \nabla_{\beta}p_{\gamma} {A}_{\gamma\tau}^{-1} p^j_{\tau}W^j 
 + p_{\gamma}\nabla_\beta {A}_{\gamma\tau}^{-1} p^j_{\tau}W^{j}\nonumber\\
& + p_\gamma {A}_{\gamma\tau}^{-1} p^j_{\tau}\nabla_{\beta}W^j
\end{align}
with $W^{j} = W (\bm{x} - \bm{x}^j)$, $\nabla_{\beta}W^{j} = \nabla_{x_\beta}W(\bm{x} - \bm{x}^j)$.
The strain tensor field, $\bm{\epsilon}(\bm{x})$, can thus be approximated by applying Eq.~\eqref{eq:gradient} to Eq.~(\eqref{eq:greenstvenant}).  Observe that $u^j_\alpha$ depend on current particle positions 
$\bm{q}$ while $\nabla_\beta \Psi_j(\bm{x})$ depend only on
the reference positions.

\subsection{Deformation constraint}
\label{sec:constraint-based}
The strain energy density, $U(\bm{x} ) = \tfrac{1}{2}\bm{\epsilon}^T \bm{C} \bm{\epsilon}$,
is transformed into a regularised constraint via a Legendre transform
as presented in Sec.~\eqref{sec:discrete_multidomain}. For each 
particle a deformation constraint is imposed
\begin{equation}
	\bm{g}^i = \begin{bmatrix}[c]
		\epsilon^i_{xx}\\
		\epsilon^i_{yy}\\
		\epsilon^i_{zz}\\
		\epsilon^i_{yz} + \epsilon^i_{zy}\\
		\epsilon^i_{zx} + \epsilon^i_{xz}\\
		\epsilon^i_{xy} + \epsilon^i_{yx}
\end{bmatrix}
\end{equation}
with particle strain components 
$\epsilon^i_{\alpha\beta} = \frac{1}{2} (s^i_{\alpha\beta} + s^i_{\beta\alpha} + s^i_{\gamma\alpha} s^i_{\gamma\beta} )$ based on the
displacement gradient $s^i_{\alpha\beta}(\bm{q}) = \nabla_\beta u_\alpha (\bm{x}^i,\bm{q})$ for particle $i$, computed by evaluating the
field at $\bm{x} = \bm{x}^i$.  The compliance parameter become
$\bm{\varepsilon}^i= (V^i\bm{C}^i)^{-1}$.  Observe
that the compliance depend only on the material parameters, 
through Eq.~(\eqref{eq:stifnesstensorreduced}), and on the
spatial resolution through the particle volume factor $V^i$ 
that appear when integrating from energy density to particle energy.

The Jacobian of the deformation constraint, $\bm{G} = \partial \bm{g} /\partial \bm{q}$ , can be expanded through the chain rule to 
\begin{equation}	
	G^{i}_{\alpha\beta}(\bm{q}) = \frac{\partial g^{i}_{\alpha}}{\partial \epsilon^i_{\gamma\eta}} 
	\frac{\partial \epsilon^i_{\gamma\eta}}{\partial s^i_{\tau\kappa}}
	\frac{\partial s^i_{\tau\kappa}}{\partial q_\beta^j}
\end{equation}
The Jacobian for constraint $i$ has block structure 
$\bm{G}^i = [\bm{G}^{i1},\hdots, \bm{G}^{ij}, \hdots, 
\bm{G}^{i\textsub{N}{p}}]$, where each block element
$\bm{G}^{ij}$ has dimension $6\times 3$. For notational
convenience the constraint vector is split in two parts, 
one for the diagonal terms of the strain tensor, 
$\textsub{\bm{g}}{d}^i = [\epsilon^i_{xx},\epsilon^i_{yy},\epsilon^i_{zz}]^T$ and 
one for the off-diagonal terms 
$\textsub{\bm{g}}{od}^i = [\epsilon^i_{yz} + \epsilon^i_{zy},\
\epsilon^i_{zx} + \epsilon^i_{xz},\ \epsilon^i_{xy} + \epsilon^i_{yx}]^T$.
The Jacobian blocks $\bm{G}^{ij}$ are splitted 
correspondingly in two $3 \times 3$ Jacobian blocks, $\bm{G}^{ij} = [{\textsub{\bm{G}}{d}^{ij}}^T,\ 
{\textsub{\bm{G}}{od}^{ij}}^T]^T$.  After some lengthy algebra the 
following expressions for the Jacobians are found
\begin{align}
\label{eq:strainjacobiansfinal}
\textsub{\bm{G}}{d}^{ij} &=  \begin{bmatrix}
(1 + s^i_{xx}) \Lambda^{ij}_x & s^i_{yx} \Lambda^{ij}_x & s^i_{zx} \Lambda^{ij}_{x}\\
s^i_{xy} \Lambda^{ij}_y & (1 + s^i_{yy}) \Lambda^{ij}_{y} & s^i_{zy} \Lambda^{ij}_{y}\\
s^i_{xz} \Lambda^{ij}_z  & s^i_{yz} \Lambda^{ij}_z  & (1 + s^i_{zz})\Lambda^{ij}_z
\end{bmatrix}\\
\textsub{\bm{G}}{od}^{ij} &= \begin{bmatrix}
s^i_{xy} \Lambda^{ij}_z + s^i_{xz} \Lambda^{ij}_y  \;&  (1 + s^i_{yy}) \Lambda^{ij}_z+s^i_{yz} \Lambda^{ij}_y \; & s^i_{zy} \Lambda^{ij}_z+(1 + s^i_{zz}) \Lambda^{ij}_y\\
s^i_{xz} \Lambda^{ij}_x + (1 + s^i_{xx}) \Lambda^{ij}_z\; & s^i_{yz} \Lambda^{ij}_x+s^i_{yx} \Lambda^{ij}_z \; & (1 + s^i_{zz}) \Lambda^j_x+s^i_{zx}\Lambda^j_z\\
(1 + s^i_{xx}) \Lambda^{ij}_y + s^i_{xy} \Lambda^{ij}_x \; & s^i_{yx} \Lambda^{ij}_y+(1 + s^i_{yy}) \Lambda^{ij}_x \; & s^i_{zx} \Lambda^{ij}_y+s^i_{zy} \Lambda^{ij}_x
\end{bmatrix}
\end{align}
\noindent
where $\Lambda^{ij}_{\beta} = \nabla_\beta \Psi_j(\bm{x}^i)$. Observe
that only $s^i_{\alpha\beta}(\bm{q})$ depend on the 
current particle positions while $\Lambda^{ij}_{\beta}$ only
depend on the reference positions and may thus be precomputed.
In the derivation of the Jacobians the following partial derivatives are used
\begin{align}
	\frac{\partial \epsilon^i_{\gamma\eta}}{\partial s^i_{\tau\kappa}} = & \tfrac{1}{2} ( 
		\delta_{\gamma\tau}^{\eta\kappa} + \delta_{\eta\tau}^{\gamma\kappa} 
		 + \delta_{\kappa\gamma}\nabla \bm{u}^i_{\tau\eta} + \delta_{\eta\kappa}\nabla \bm{u}^i_{\tau\gamma})\nonumber\\
	\frac{\partial s^i_{\tau\kappa}}{\partial q_\beta^j} = & \delta_{\tau\beta} \nabla_\kappa \Psi_j(\bm{x}^i)\nonumber
\end{align}
with Kronecker notation $\delta_{\gamma\tau}^{\eta\kappa}\equiv \delta_{\gamma\tau}\delta_{\eta\kappa}$.
\section{Simulation}
\label{sec:simulation}
The procedure, implementation details and results for
nonsmooth multidomain dynamic simulation including elastoplastic solids 
is presented in this section.  The main simulation algorithm is given
in section \ref{sec:mainsteps}. The elastoplastic solid model is implemented in 
C\texttt{++} as an extension to the simulation software \textit{AgX Dynamics} \cite{agx}
making use of its data types, collision detection algorithms and solver framework, 
which support simulation of nonsmooth rigid multibody dynamics systems in the formulation of 
section \ref{sec:discrete_multidomain}.
The \textsc{mlcp} solver framework is covered in section 
\ref{sec:mlcpsolver}. The elastic and plastic model 
and implementation is verified by performing load-displacement 
simulation tests presented in section \ref{sec:verification}. Finally,
in section \ref{sec:multidomainex}, the applicability of the approach is 
demonstrated by multidomain examples including a simple terrain vehicle 
with tracked bogies driving over a deformable terrain and simulated
cone penetrometer test in soil with embedded rock.

\subsection{Main algorithm}
\label{sec:mainsteps}
Algorithm \ref{alg:mainalg} lists the main steps in 
running a multidomain dynamics simulation including both elastoplastic 
solids and contacting rigid multibodies.

\begin{algorithm}
  \caption{Main simulation algorithm}
  \label{alg:mainalg}
  \begin{algorithmic}[1]
  \STATE initialisation of bodies and constraints
  \STATE compute \textsc{mlcp} blocks $\bm{M}, \bm{\Sigma}, \bm{\Upsilon}$
  \STATE compute \textsc{mls} blocks $\bm{A}^{-1}_i$, $\bm{\Lambda}_i$
  \FORALL{timestep $n = 1,2,\hdots $}
  \STATE get input signals and explicit forces
  \STATE do contact detection
  \FORALL{particles $i = 1,2,\hdots \textsub{N}{p}$}
  \STATE compute $\bm{u}^i_{n}$ and $\bm{\epsilon}^i_n = {\bm{J}^i_n}^T\bm{J}^i_n - \bm{1}$
  \STATE $[\bm{\epsilon}^{\text{e},i}_n, \bm{\epsilon}^{\text{p},i}_n] = \mathrm{radial\_return}(\bm{\epsilon}^i_n,\bm{\epsilon}^{\text{p},i}_{n-1})$ 
  \ENDFOR
  \STATE compute constraint data $\bm{G},\bm{g}$ 
  \STATE build \textsc{mlcp} data $\bm{H},\bm{r},\bm{l},\bm{u}$
  \STATE solve $\bm{z} = $ \textsc{mlcp(}$\bm{H},\bm{r},\bm{l},\bm{u}$\textsc{)}
  \STATE get $\dot{\bm{q}}_{n+1}$ and constraint force $\bm{G}^T\bm{\lambda}$
  \STATE update position $\bm{q}_{n+1} = \bm{q}_{n} + h \dot{\bm{q}}_{n+1}$
  \STATE store and visualise the new state
  \ENDFOR
  \STATE post-process and visualise stored data
  \end{algorithmic}
\end{algorithm}

\begin{algorithm}
  \caption{Plastic radial return algorithm}
  \label{alg:radial_return}
  \begin{algorithmic}[1]
  \STATE input total strain $\bm{\epsilon}_{n}$ and stored plastic $\bm{\epsilon}^{\text{p}}_{n-1}$
  \STATE trial plastic strain $\bm{\epsilon}^{\text{p}}_{n} = \bm{\epsilon}^{\text{p}}_{n-1}$
    \STATE compute trial stress $\bm{\sigma}_{n} = \bm{C}(\bm{\epsilon}_{n} - \bm{\epsilon}
    _{n}^{\text{p}})$  
    \IF{yielding $\Phi(\bm{\sigma}_{n}) > 0$}
    \STATE return to surface 
    \WHILE{$\left| \Phi(\bm{\sigma}_{n})\right| > \textsub{\varepsilon}{tol}$} 
    \STATE{$\Delta \bm{\lambda} = \frac{\Phi(\bm{\sigma}_{n})}{\frac{\partial \Phi}{\partial 
    \bm{\sigma}}\textsub{\bm{C}}{}\frac{\partial \Psi}{\partial \bm{\sigma}}}$}     
    \STATE{$\Delta \bm{\epsilon}^{\text{p}} = \Delta \bm{\lambda} \frac{\partial \Psi}
    {\partial \bm{\sigma}}$}     
    \STATE{$\Delta \bm{\sigma} =  \Delta \bm{\lambda} \bm{C}\frac{\partial \Psi}{\partial 
    \bm{\sigma}}$}     
    \STATE{$\bm{\epsilon}^{\text{p}}_{n} = \bm{\epsilon}^{\text{p}}_{n} + \Delta 
    \bm{\epsilon}^{\text{p}}$}
    \STATE{$\bm{\sigma}_{n} = \bm{\sigma}_{n} - \Delta \bm{\sigma}$}
    \ENDWHILE  
    \STATE{update cap variables $\kappa_n(\bm{\epsilon}^{\text{p}})$ and ${I_1^c}_{n}(\bm{\sigma}_{n})$} 
    \ENDIF
    \STATE compute elastic strain $\bm{\epsilon}^{\text{e}}_{n} = \bm{C}^{-1} \bm{\sigma}
    _{n}$
  \end{algorithmic}
\end{algorithm}

If the trial stress is outside the elastic domain
the radial return algorithm \ref{alg:radial_return} return
the system radially to the nearest point on the yield surface, 
either to the Drucker-Prager surface, or to the tension or compression cap, 
see Fig.~\ref{fig:gapfunction}.  After returning the stress to the yield surface,
with an error threshold $\textsub{\varepsilon}{tol}$, the compaction
variable $\kappa$ is updated.

The elastoplastic solid is initialised by defining a solid volume and
assigning material parameters.  The solid is discretised into $\textsub{N}{p}$ pseudo-particles
according to a given spatial resolution.  For simplicity, the particles are positioned in
a regular cubic grid.  This defines the reference state with vanishing strain and stress.  Any
initial displacement may be applied to the particles.  The elastoplastic constraints
are initialised and connectivity data listing the particles involved in each constraint.
Similarly, rigid bodies and kinematic constraints are defined, 
for instance, to form an articulated vehicle and powertrain.
Each body is assigned a geometric 3D shape.  
Contact properties are assigned to each body,
including coefficient of restitution, elasticity and friction.
These are used to generate contact constraint data included in the \textsc{mlcp}
when triggered by contact detection algorithm.
The pseudo-particles are given a spherical geometric shape
for dynamic contacts with rigid bodies and static geometries. 
Fixed-point or plane constraints can also be defined to
model permanent boundary conditions.
Contacts between particles are disabled.
Certain quantities are computed once before time integration and then reused for the sake
of optimisation, e.g., the inverse moment matrix $\bm{A}^{-1}$ 
in the \textsc{mls} approximation, $\nabla_\beta\Phi$ and $\bm{\Lambda}$.

The simulation is run with fixed timestep, following the stepper in section 
\ref{sec:time_discretisation}.
Each timestep begin with reading input signals, e.g., from operator and control system,
and computation of explicit forces.  Next, contact detection algorithms 
produce a set of contacts for intersecting geometries.  
Each contact position and velocity, penetration depth, normal 
and tangent is stored.  Contacts are classified as either impacting,
continuous or separating, depending on the sign of the relative
contact velocity.

The strain and stress fields are computed as described in section \ref{sec:elastoplastic}
using the \textsc{mls} approximation of the displacement field for the spatial discretisation described in section 
\ref{sec:meshfree}. Plastic deformations are handled by the return algorithm and capped Drucker-Prager model as listed in algorithm \ref{alg:radial_return}.
Constraint violation and Jacobians are computed from elastic strain, contact data 
and state of the multibodies. The matrix
blocks and limits for the \textsc{mlcp} are computed and fed to the \textsc{mlcp} 
solver outlined in section \ref{sec:mlcpsolver}.  For highly elastic contacts
the main solve is preceded with a \emph{impact solve stage}, in which
impact contacts are solved using the Newton impulse law 
$\bm{G}\dot{\bm{q}}^{+} = - e \bm{G}\dot{\bm{q}}^{-}$, with
coefficient of restitution $e$, while maintaining all other constraint
velocities satisfied, $\bm{G}\dot{\bm{q}} = 0$.  The main \textsc{mlcp} solve
produces the new velocities and constraint forces, which are extracted and 
stored.  After this the positions and orientations are integrated.
At the end of each timestep data of the new state is stored for 
post-processing and visualisation. 

\subsection{Overview of the \textsc{mlcp} solver}
\label{sec:mlcpsolver}

The main computational task each timestep is solving
the \textsc{mlcp} in equation \eqref{eq:mlcp} with saddle-point
matrix structure given in equation \eqref{eq:saddle_point}.
Each sub-matrix is block-sparse and the routines for building, storing and
solving are tailored for this and exploit \textsc{blas3} operations as much as 
possible to gain computational speed.
In order to achieve high performance, e.g., realtime simulation of complex 
multidomain systems, splitting is applied \cite{cottle:1992:lcp}
such that subproblems can be treated with different \textsc{mlcp} 
solvers that best fit the requirements of accuracy, stability
and scalability.  For the elastoplastic solid and for 
jointed rigid multibodies, with relatively few complementarity conditions,
a direct \textsc{mlcp} solver is used that is described in section
\ref{sec:direct}.  

\subsubsection{Split solve}
Assuming three sets of constraints, labelled $A$, $B$ and $AB$
for subsystem $A$ and $B$ and their coupling.
The linear system \eqref{eq:self_similar} is split into subproblems 
by duplicating the variables $\dot{\bm{q}}$ and $\bm{\lambda}_{AB}$ and reordering 
the augmented system.  A stationary iterative update
procedure can then be formed based on the four matrix blocks 
\begin{align}
  \label{eq:split11}
  \bm{H}_{AB} \bm{z}^{k+1}_{AB} & = - \bm{r}_{AB} + \bm{G}^T_{B}\bm{\lambda}^{k}_{B} \\
  \label{eq:split22}
  \bm{H}_{BA} \bm{z}^{k+1}_{BA} & = - \bm{r}_{BA} + \bm{G}^T_{BA}\bm{\lambda}^{k+1}_{A}
\end{align}
with
\begin{equation}
	\bm{H}_{AB} = 
  \begin{bmatrix}[ccc] 
  \bm{M} & -\bm{G}^T_{A} & -\bm{G}^T_{AB}\\ 
  \bm{G}^T_{A} & \bm{\Sigma}_{A} & 0\\
  \bm{G}^T_{AB} & 0 & \bm{\Sigma}_{AB}\\
  \end{bmatrix}
  \label{eq:H_AB}
\end{equation}
\begin{equation}
	\bm{z}_{AB}  =   \begin{bmatrix} \dot{\bm{q}} \\ \bm{\lambda}_A \\ \bm{\lambda}_{AB} \end{bmatrix}\ ,\ \ 
	\bm{r}_{AB}  =   \begin{bmatrix} \dot{\bm{p}} \\ \bm{v}_A \\ \bm{v}_{AB} \end{bmatrix}
  \label{eq:z_AB}
\end{equation}
and the same with $A$ and $B$ permuted.  The iterative
procedure converges if the spectral radius of the augmented system
$\bm{H}$ fulfils $\rho([\bm{D} + \bm{L}]^{-1}\bm{U}) < 1$. 
The key point is that each subproblem can be approached using different solvers.
In the case of a machine interacting with an elastoplastic terrain, the 
machine and solid terrain constraints, $A$, are split from the tangential contact 
forces related to dry friction, $B$, both systems sharing the normal contact 
constraints (nonpenetration), $AB$.  
The subsystem \eqref{eq:split11} with the solid, machine and contact normals is 
solved first using a direct solver, while
the subsystem \eqref{eq:split22} containing friction 
constraints is solved using an iterative projected Gauss-Seidel solver.
The split solve may be terminated after the first iteration, accepting errors from 
the friction forces, or continued with a final stage 1 solve.
Experiments show that further iterations do not necessarily reduce 
the errors.  For large contact 
systems ($>$1K contacts), with many complementarity conditions, 
both normal and friction constraints may be moved to an iterative 
projected Gauss-Seidel solver \cite{servin:2014:esn}.

\subsubsection{Direct solve}
\label{sec:direct}

The direct solver is a direct block pivot method \cite{cottle:1992:lcp} 
for \textsc{mlcp}s based on Newton-Rhapson iterations applied to nonsmooth formulation.
The detailed algorithm is an adaptation of Murty's principal pivoting method, 
and is found in Ref.~\cite{lacoursiere:2013:svt} as Algorithm 21.2.  
Before the solve step, the matrix $\bm{H}$ is factorised by a permuted 
\textsc{ldl}-factorisation, $\bm{H} =\bm{P}\bm{L}\bm{D}\bm{L}^T \bm{P}^T$, 
where the permutation $\bm{P}$ is used to minimise fill-ins and may
include leaf-swapping \cite{leafswopping_lacoursiere_linde} to 
avoid pivoting on $\bm{\Sigma}$ since it is often close to zero.
The factorisation requires symmetric indefinite matrices but this is not 
the case with $\bm{H}$ initially as seen in equation \eqref{eq:self_similar}. 
This is fixed by absorbing a negative sign in the vector $\bm{z}$.

\subsection{Verification tests}
\label{sec:verification}

\subsubsection{Elasticity}
\label{sec:elastic_verification}
The validity of the elastic solid model is confirmed by a number of tests of 
macroscopic relations between an applied load and the resulting displacement. 

\begin{figure}[!ht]
\centering
\subfigure{
\includegraphics[width=0.17\textwidth]{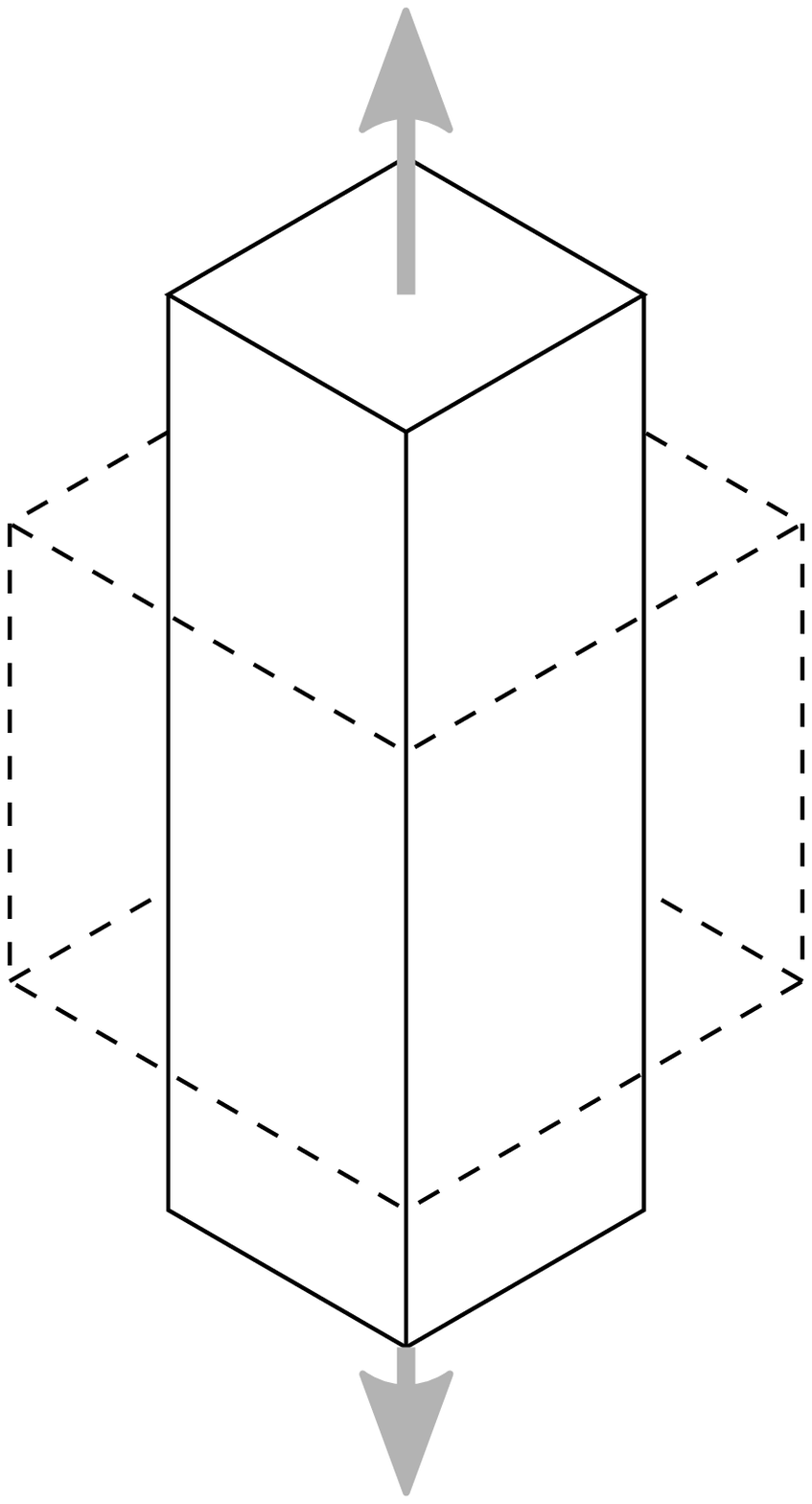}}\qquad
\subfigure{
\includegraphics[width=0.23\textwidth]{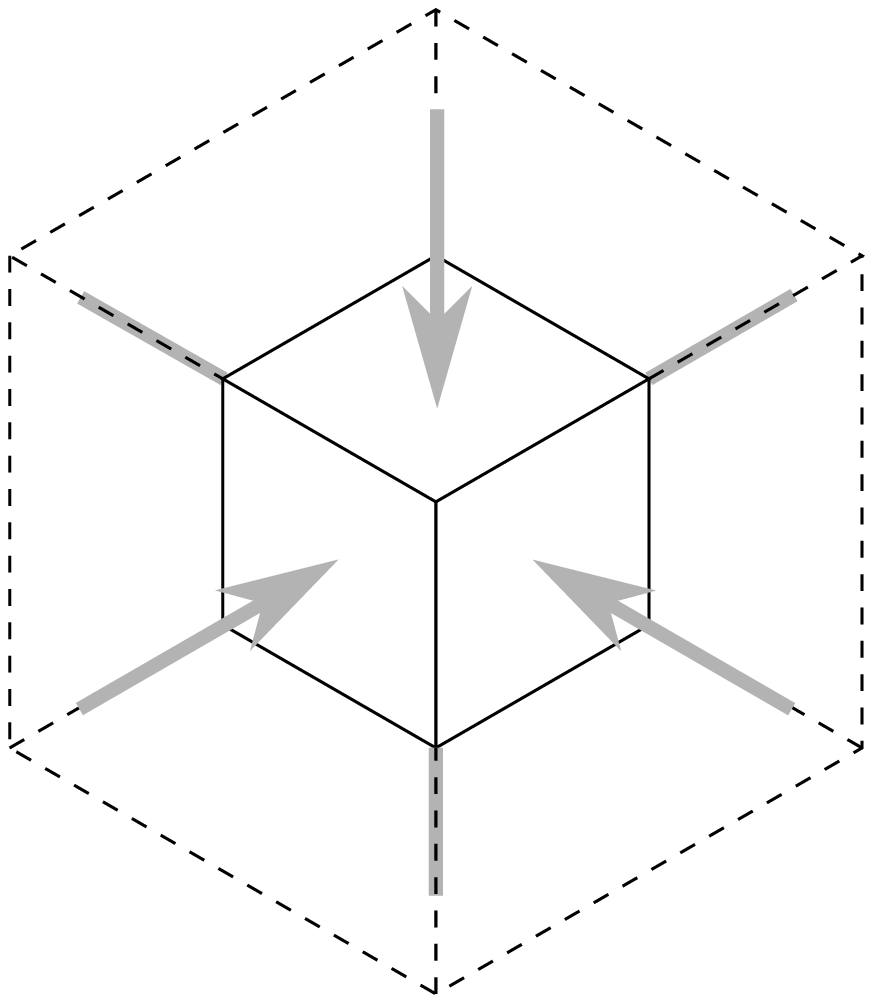}}
\caption{Illustration of the verification tests by 
uniaxial stretch (left) and hydrostatic compression (right)
of an elastic cube.}
\label{fig:tests}
\end{figure}

Uniaxial stretching and hydrostatic compression are investigated, as 
described in figure \ref{fig:tests}.

\begin{figure}
\centering
\includegraphics[width=0.45\textwidth]{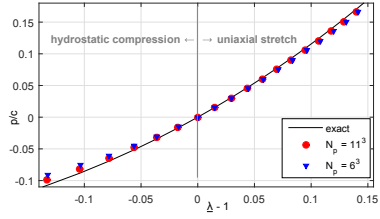}
\caption{Verification test of elastic response to hydrostatic compression ($\underline{\lambda} < 1$) and uniaxial stretch ($\underline{\lambda} > 1$) with $E=10^6$ Pa, $\nu=0.25$ and $\textsub{N}{p}=11^3$ (circles) and $\textsub{N}{p}=6^3$ (triangles).  The analytical solution is represented with a solid curve. }
\label{fig:elastic_verification}
\end{figure}

In the uniaxial test the boundary condition in the tensile direction
is a plane constraint with free slip in plane.  The other boundaries are free. 
In the hydrostatic compression test the boundary conditions are dynamically
generated unilateral contact constraints.  The boundary force is increased linearly with time,
by regulating the position of the boundary geometry, while 
measuring the stress and strain components of equations
\eqref{eq:error_trueuniaxial}-\eqref{eq:error_truebulk}.
All tests are simulated on a uniformly discretised homogeneous isotropic three 
dimensional solid with varying parameters as described in table 
\ref{tab:verification}.  
\begin{table}
  \centering
  \begin{tabular}{lr}
    Parameter & Value\\\hline
    side length & $1$ m \\
    $\textsub{N}{p}$ & $\left( 6^3, 9^3, 11^3 \right)$\\
    resolution & $1/\textsub{N}{p}^{3}$ m \\
    influence radii & $2$ $\times$ resolution\\
    mass density & $2000$ kg/m$^3$\\
    Young's modulus $E$ & $\left( 10^6,10^8 \right) $ Pa\\
    Poisson ratio $\nu$ & $\left( 0.1,0.25,0.49 \right)$\\
    Timestep $h$ & $10^{-3}$ s\\\hline
  \end{tabular}
  \caption{Elastic verification parameters}
  \label{tab:verification}
\end{table}
The simulation result is compared with elasticity theory that for
St.~Venant-Kirchoff materials imply
the following exact relations between applied load and uniform deformation 
\begin{align}
  \label{eq:error_trueuniaxial}
    \textsub{p}{u} &=  \frac{1}{2} \textsub{c}{u} \textsub{\underline{\lambda}}{u} \left( \textsub{\underline{\lambda}}{u}^2 - 1 \right)\\
  \label{eq:error_truebulk}
    \textsub{p}{c} &= \frac{1}{2} \textsub{c}{c} \textsub{\underline{\lambda}}{c}\left( \textsub{\underline{\lambda}}{c}^2 - 1 \right) 
\end{align}
where $p = F / A_0$ is the load pressure on the boundary, $\textsub{c}{u} = E$
$\textsub{c}{c} = \frac{E}{(1 - 2\nu)}$ are the Young's and bulk modulus,
and $\textsub{\underline{\lambda}}{u}$ and $\textsub{\underline{\lambda}}{c}$ 
are the stretch ratios $l/l_0$ of the test-cube with the deformed side-length $l$, initial rest length $l_0$ and cross-sectional area $A_0$.  The simulation results for $E = 10^6$ Pa and
$\nu = 0.25$ are presented in figure \ref{fig:elastic_verification}.  

For large deformations ($15$ \%) the 
simulated result agrees with the exact solution to an accuracy of $5$ \% 
for uniaxial stretch and $10$ \% for hydrostatic compression.
The error decrease with size of the deformation and with finer resolution.
Observe that for infinitesimal deformations 
$\frac{1}{2}\underline{\lambda} \left( \underline{\lambda}^2 - 1\right) \approx \Delta l/l_0$, Eq.~
\eqref{eq:error_trueuniaxial}-\eqref{eq:error_truebulk} approximates to the well-known expressions from 
Hooke's law.  Increasing the stiffness or resolution further require smaller time-step
for numerical stability.  

\begin{figure}
\centering
\includegraphics[width=0.49\textwidth]{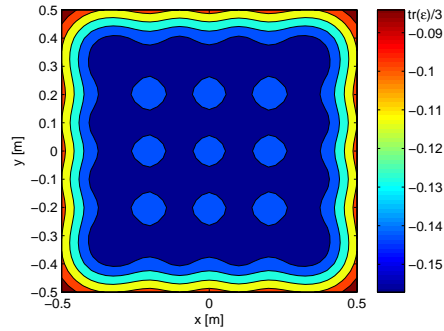}
\caption{The variation of $\tr(\epsilon)$ through the cross-section of the
elastic test cube under hydrostatic compression to $\underline{\lambda} = 0.86$.}
\label{fig:elastic_hydrostatic_error}
\end{figure}

\begin{figure}
\centering
\includegraphics[width=0.49\textwidth]{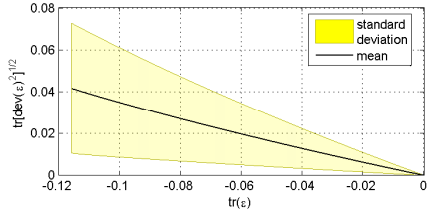}
\caption{The variation of $\tr[\text{dev}(\epsilon)^2]^{1/2}$ 
during hydrostatic compression.}
\label{fig:strain_deviator}
\end{figure}

In order to understand the nature of the error the strain field under
hydrostatic compression is analysed, see figure \ref{fig:elastic_hydrostatic_error}.
The strain field is not uniform through the material as expected but deviate
as much as $30$ \% and most near the boundary.  Enforcing a uniform deformation,
by initial particle displacements, produce a perfectly uniform strain and stress
field.  We thus conclude that the effect is caused by the way that boundary 
conditions are enforced.  See, section \ref{sec:conclusions} for a further
discussion on this.  
The errors in the strain distribution propagate to the stress and ultimately
to the plastic behaviour also.  In particular the deviation from 
nonuniform strain will lead to a non-vanishing stress deviator in deformations
where it is expected to be zero.  The mean strain deviatior and its
standard deviation in the case of hydrostatic compression is displayed in figure \ref{fig:strain_deviator}.  

\subsubsection{Plasticity}
\label{sec:elastoplastic_verification}
\begin{figure}
\centering
\subfigure[Stress evolution.]{\includegraphics[width=0.49\textwidth]{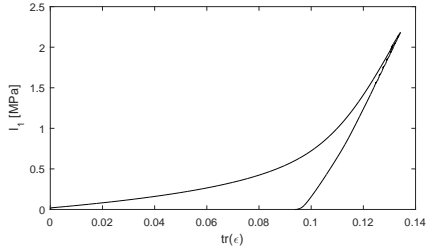}}
\subfigure[Mean stress evolution and yield surface.]{\includegraphics[width=0.49\textwidth]{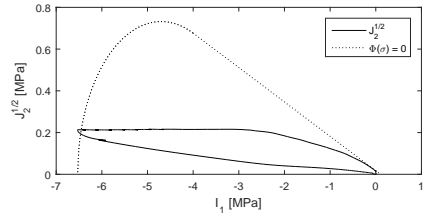}}
\caption{Hydrostatic compression and unloading of an elastoplastic solid cube.}
\label{fig:hydrostatic_plastic}
\end{figure}

The Drucker-Prager cap model is tested under hydrostatic and triaxial loading and 
unloading, following the tests made in \cite{Dolarevic:2007} with 
$E = 9.0865 \cdot 10^7$ Pa,$\nu=0.2981$, $c = 14$ kPa, $\phi = 22^\circ$, $W = 0.1$,
$D = 1$ mm$^2$/N.  The first test examines the relation 
between stress and strain while undergoing plastic deformation on the compression cap,
see figure \ref{fig:hydrostatic_plastic}.  

The test produce a permanent deformation of the cube by 
$\tr{\epsilon} = 0.095$ compared to $\tr{\epsilon} = 0.12$ in \cite{Dolarevic:2007}.
As can be expected this deviation is due to the deviation of uniform strain,
observed in section \ref{sec:elastic_verification}, that lead to non-vanishing
deviatoric stress.  Investigating the evolution of the stress invariants, 
$I_1$ and $J_2$, in the yield space, it is clear that the compression cap is
not reached precisely on the hydrostatic axis and thus yield for a smaller of $I_1$.
The plastic shear is also not uniform which affect the permanent deformation.

\begin{figure}
\centering
\subfigure[Applied stress over deformation.]{\includegraphics[width=0.49\textwidth]{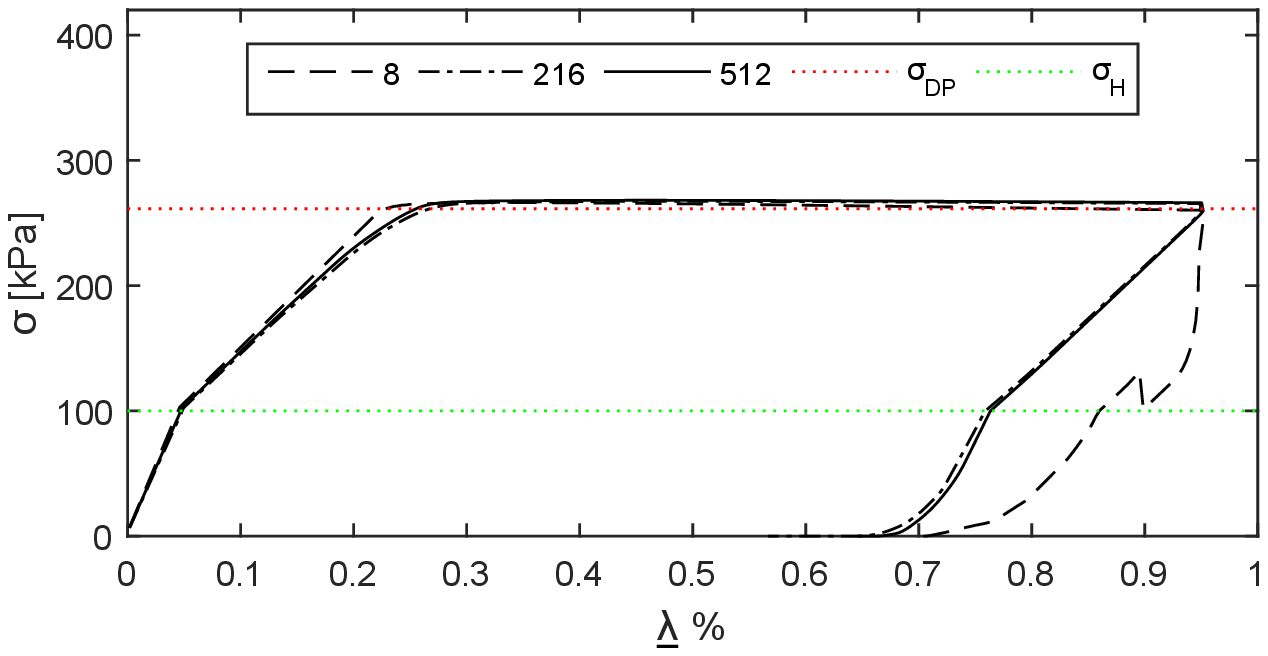}}
\subfigure[Mean stress evolution and yield surface.]{\includegraphics[width=0.49\textwidth]{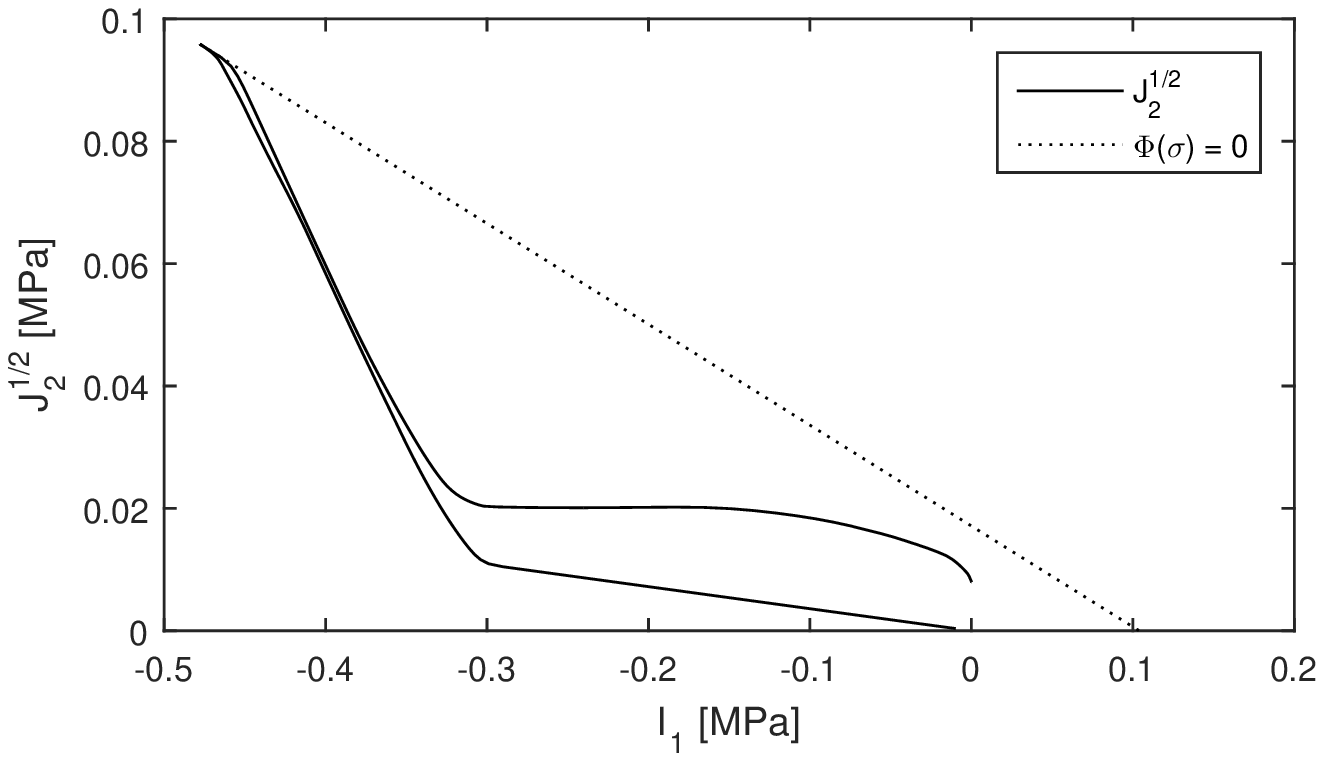}}
\caption{Plastic Drucker-Prager triaxial load and unload.}
\label{fig:triaxial_plastic_dp}
\end{figure}

\begin{figure}
\centering
\subfigure[Applied stress over deformation.]{\includegraphics[width=0.49\textwidth]{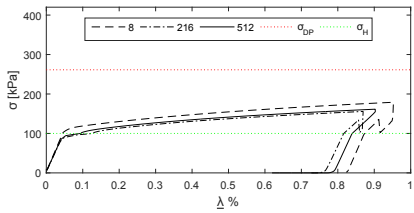}}
\subfigure[Mean stress evolution and yield surface.]{\includegraphics[width=0.49\textwidth]{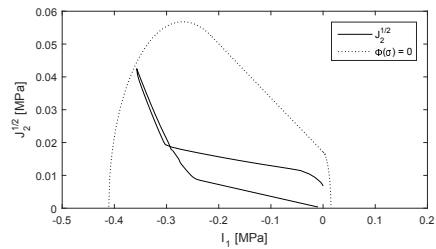}}
\caption{Plastic Cap/Drucker-Prager triaxial load and unload.}
\label{fig:triaxial_plastic}
\end{figure}

In the triaxial test, a hydrostatic pressure of $100$ kPa is first established.  The load along one axis is 
then increased up to a level where the material yields and while pressure is kept fix.  The
material is finally unloaded.  The simulated evolution of the deviatoric stress as function of 
strain is displayed in figure \ref{fig:triaxial_plastic_dp} for a Drucker-Prager material
and in figure \ref{fig:triaxial_plastic} for a capped Drucker-Prager material.  
The simulated Drucker-Prager material is found to yield at a critical 
stress of $260$ kPa.  This is in precise agreement with the analytical 
prediction from equation (\ref{eq:DP}) with hydrostatic pressure of $100$ kPa.
For the capped Drucker-Prager the yield plateau is lower and less pronounced.  
The difference is expected since the stress reach the compression cap 
surface before the Drucker-Prager surface.  The result show a weak 
dependency on resolution $\textsub{N}{p} = 2^3, 6^3$ and $8^3$. 
The triaxial test can be used to
determine the value of the plastic hardening parameter $D$, whereas the
maximum compaction parameter, $W$, can be determined from the 
hydrostatic compression test.  

\subsubsection{Dynamic contacts}
\begin{figure}
\centering
\includegraphics[width=0.45\textwidth]{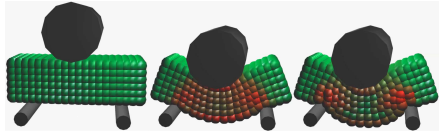}
\caption{Deformation of an elastic and elastoplastic beam (middle and right from 
initial state to the left).
The color codes
the frobenius norm of the strain tensor, ranging from $0$ (green) to $0.20$ (red).}
\label{fig:beam}
\end{figure}

Dynamic contact handling is demonstrated by dropping a beam 
to rest on two thin cylinders and then let if deformed by pressing 
a larger cylinder down on the beam.  The result for an elastic and
elastoplastic beam are displayed in Figure \ref{fig:beam}.

\subsection{Multidomain demonstration}
\label{sec:multidomainex}
\begin{figure}
\centering
\subfigure{\includegraphics[width=0.45\textwidth]{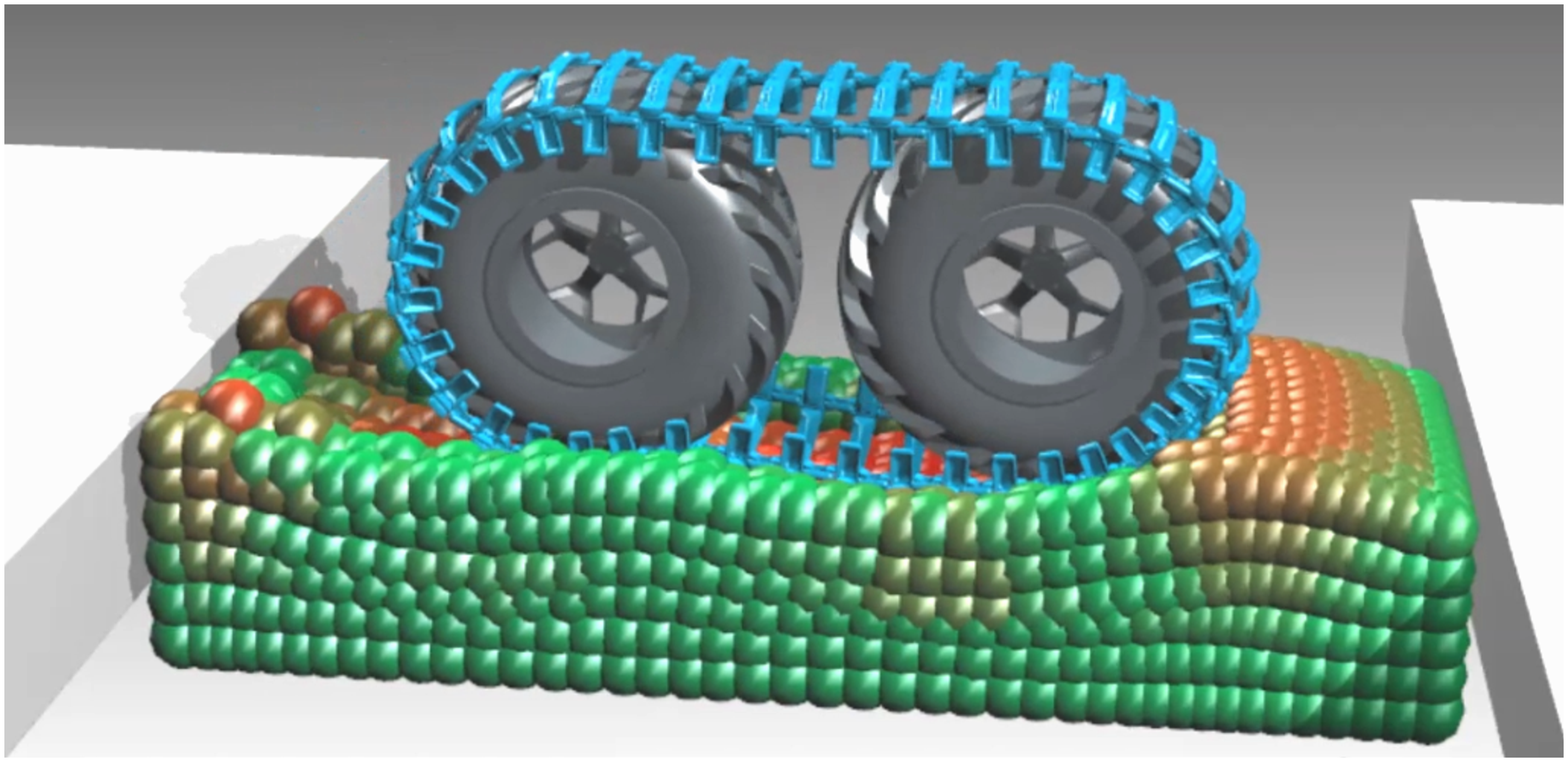}}
\subfigure{\includegraphics[width=0.45\textwidth]{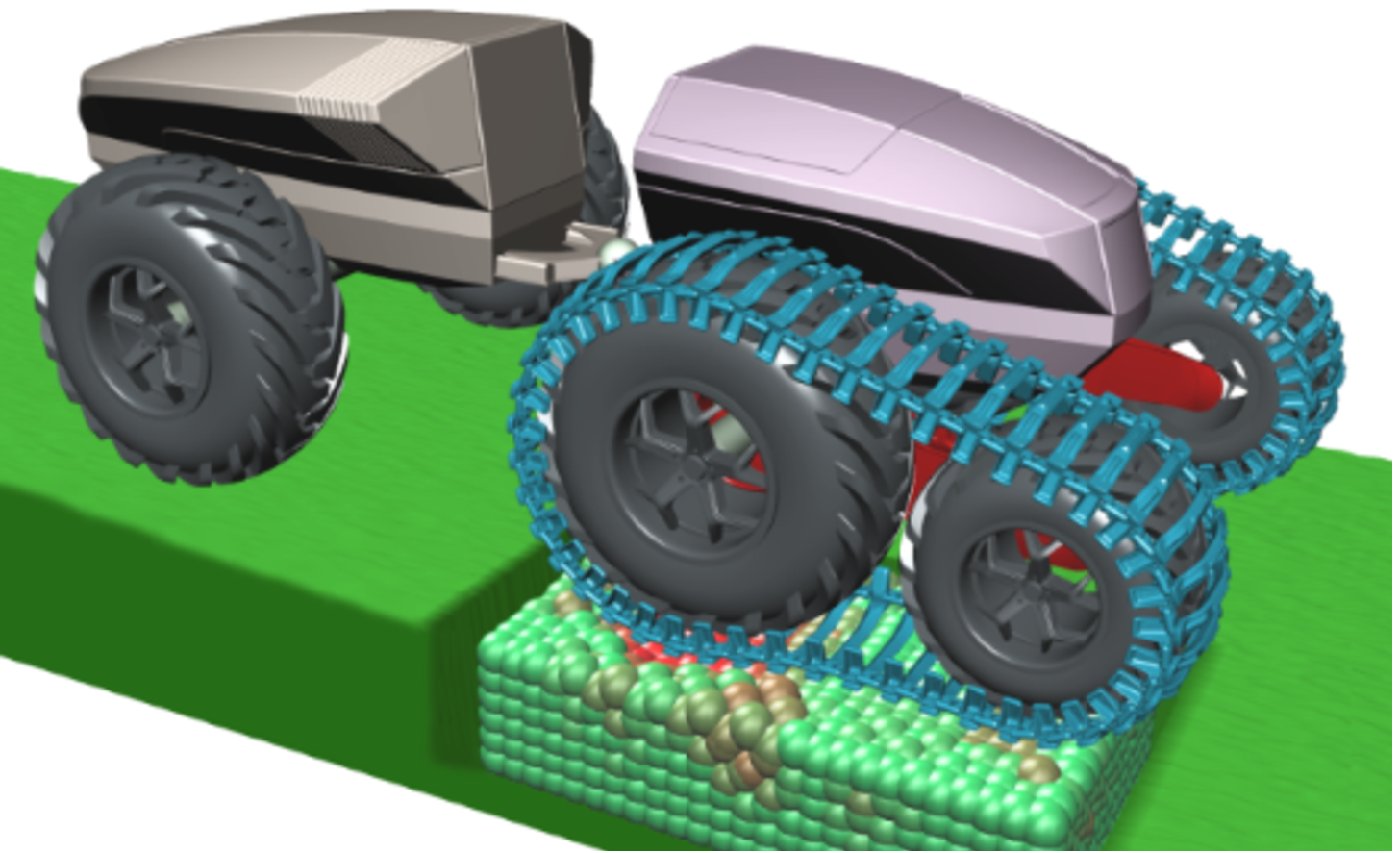}}
\caption{Images from simulation of a tracked bogie (top) and a
full terrain vehicle (bottom) passing over a zone with elastoplastic material.}
\label{fig:demo}
\end{figure}

In order to demonstrate multidomain capability with the method, two
example systems are simulated. The first system is an articulated
terrain vehicle with tracked bogies driving over a deformable 
terrain. Images from simulation of a bogie and of a full vehicle 
are shown in figure \ref{fig:demo}.
Videos from simulations are available as supplementary material
at \texttt{http://umit.cs.umu.se/elastoplastic/}.
The vehicle weighs 4 ton and consist of roughly 200 rigid bodies 
including a front and rear chassis, 
bogie frames, wheels and track elements.
These are interconnected with
kinematic constraints to model chassis articulation, bogie and wheel axes,
and tracks covering the wheeled bogies.  The vehicle drivetrain consist
of a engine, torque converter, drive shafts and differentials that 
distributes the engine power to front and rear part 
and further between left and right bogies to the wheels.
The terrain is modeled as a static trimesh and a rectangular ditch 
with elastoplastic material.
The elastoplastic material parameters are set to $Y=1$ MPa, $v=0.25$, $\phi=22^\circ$, $c = 1$ kPa,
$R_C = 4$, $T=5$ kPa, $\kappa_0=-0.5$ MPa, $D=10^{-6}$, $W = 0.2$, $\rho=2000$ kg/m$^3$, which represents a weak and soft forest terrain.   The tracked bogie and vehicle 
create a rutting with permanent deformation and the rut depth can be measured.
In the full vehicle simulation the solid is discretised into $\textsub{N}{p} = 2100$ pseudo-particles.  The coupled system of vehicle
and terrain thus have roughly 7000 degrees of freedom and  
roughly the same number of constraint equations.  With $1$ ms timestep
the simulation was of the order $1000$ slower than realtime on a 
conventional desktop computer and using a single core.  
It should be emphasised that this measure is based on prototype 
code with little effort on optimizing it.  Also the use of a direct solver 
for the terrain can be questioned.  Both the model
uncertainty and spatial discretisation error are many orders in magnitude beyond the 
accuracy delivered by the solver. %

\begin{figure}
\centering
\subfigure[$0s$]{\includegraphics[width=0.24\textwidth,clip,trim=400 0 400 0]{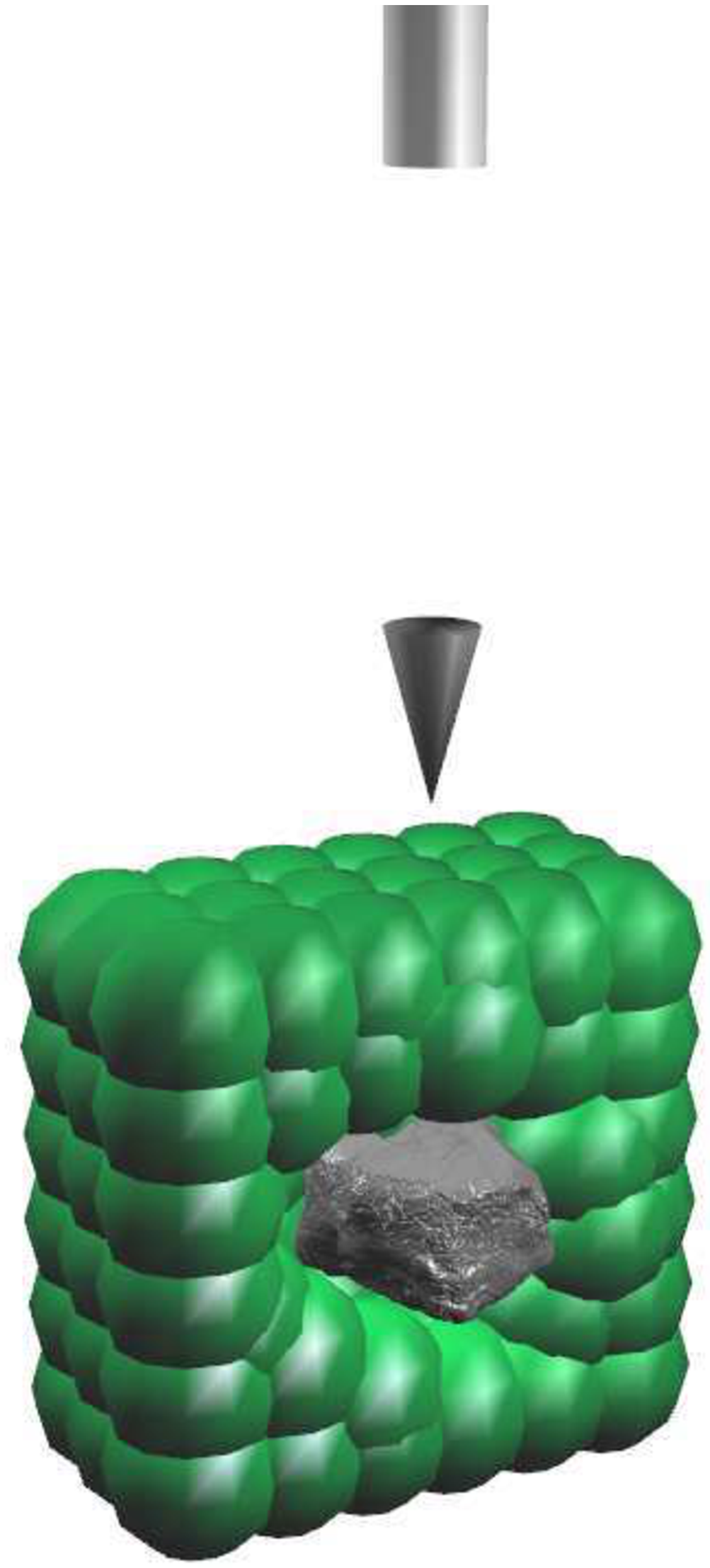}}
\subfigure[$2.4s$]{\includegraphics[width=0.24\textwidth,clip,trim=400 0 400 0]{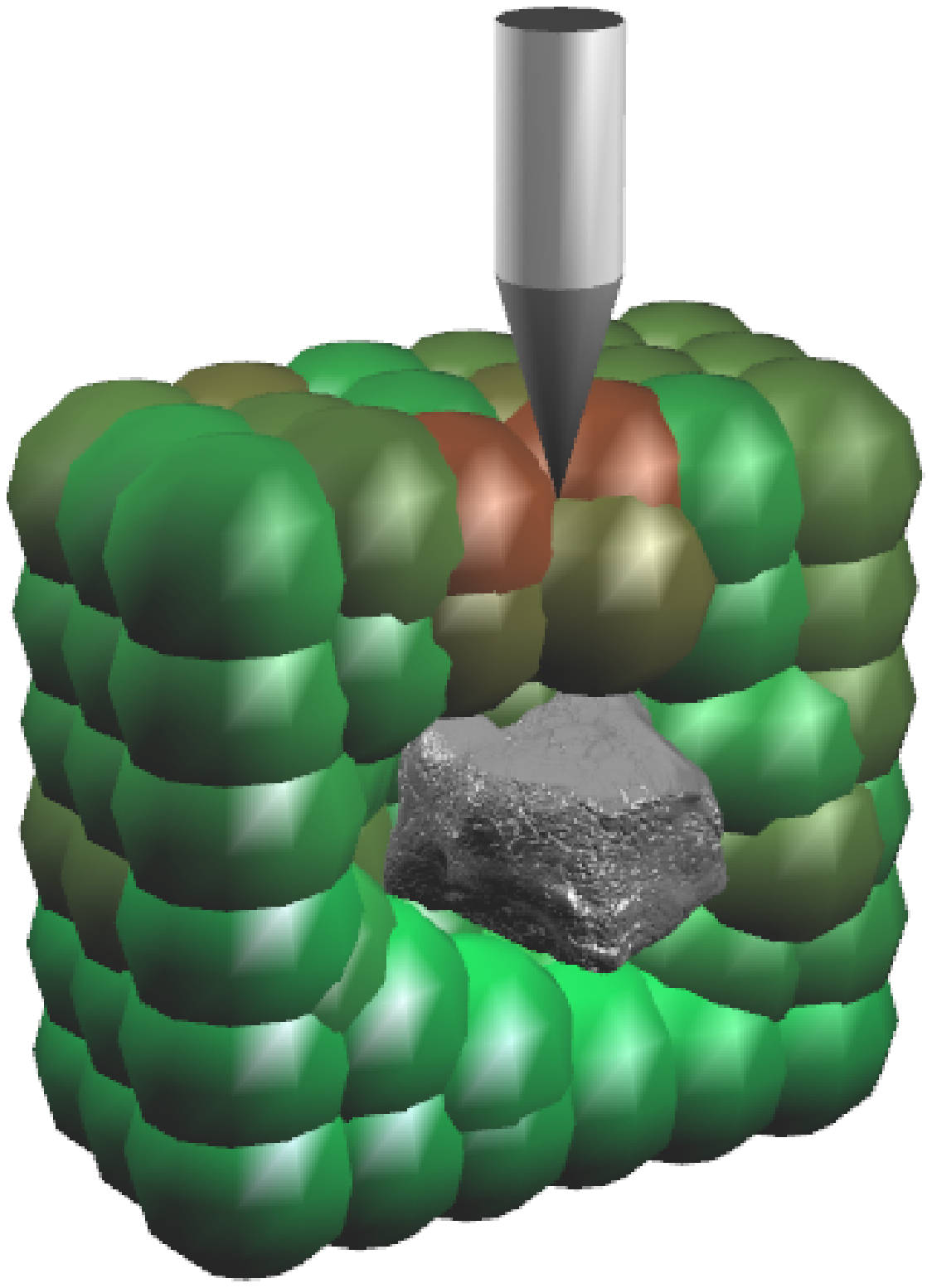}}
\subfigure[$4.7s$]{\includegraphics[width=0.24\textwidth,clip,trim=400 0 400 300]{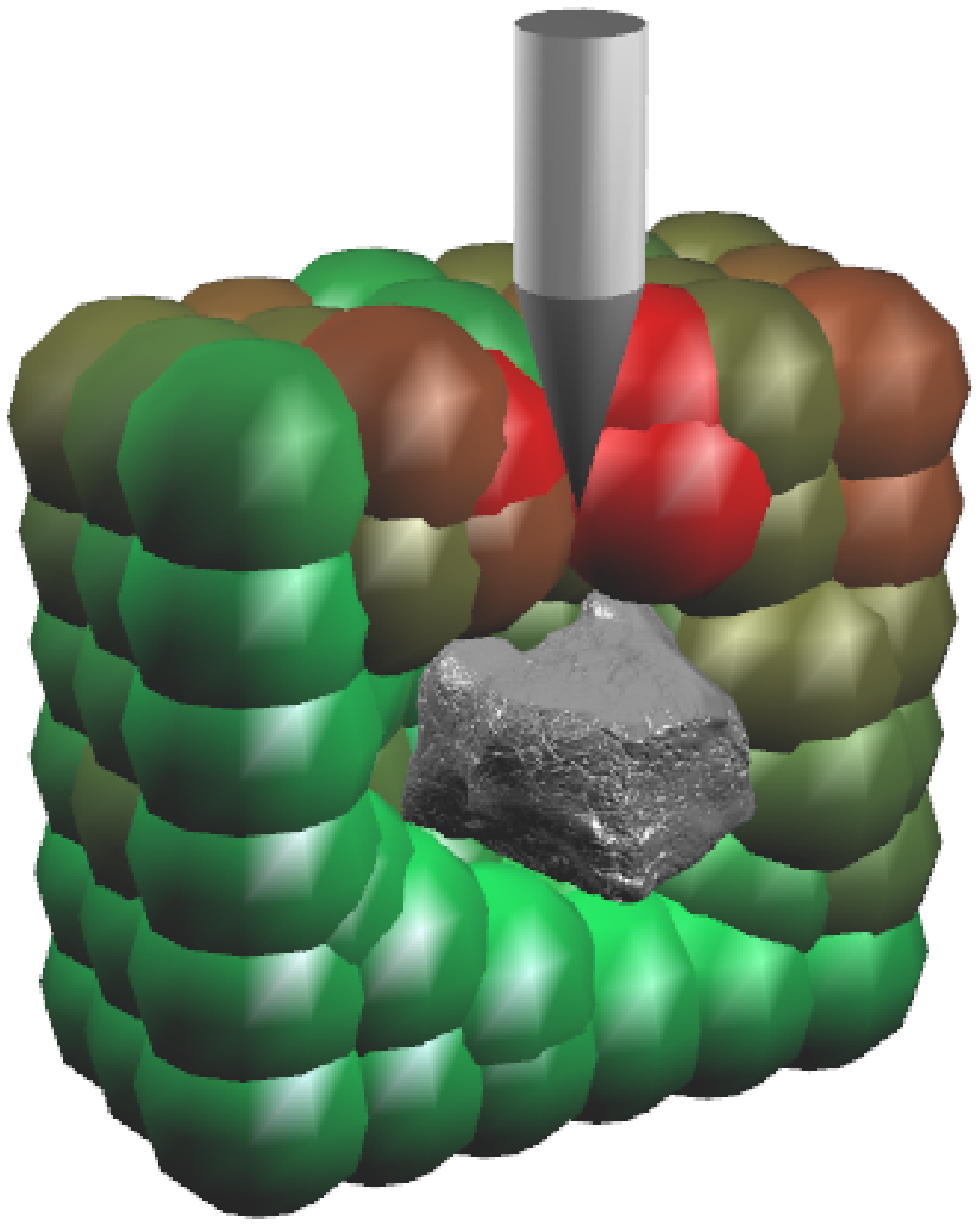}}
\subfigure[$7.0s$]{\includegraphics[width=0.24\textwidth,clip,trim=400 0 400 300]{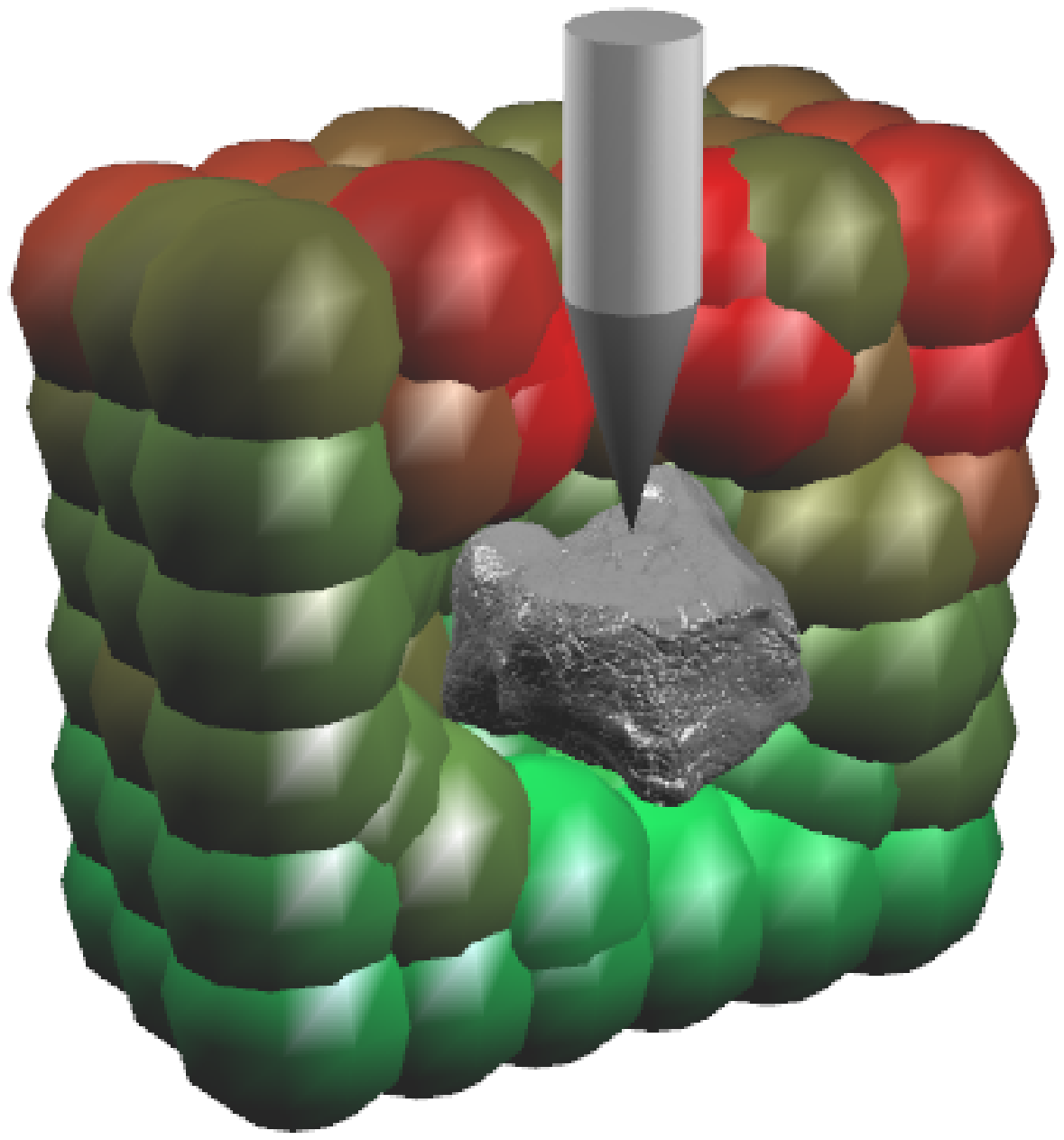}}
\caption{Image sequence from a simulated cone 
penetration test on an elastoplastic terrain with an
embedded rock.}
\label{fig:plastic_cone}
\end{figure}

\begin{figure}
\centering
\subfigure[Homogeneous terrain.]{\includegraphics[width=0.49\textwidth]{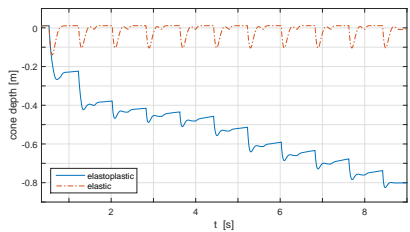}}
\subfigure[Terrain with embedded rock. The $\times$ indicate the time instances in Figure~\ref{fig:plastic_cone} b-d.]{\includegraphics[width=0.49\textwidth]{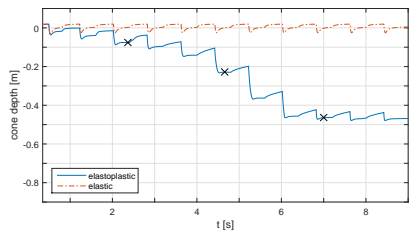}}
\caption{Simulation measurement of a cone penetrometer on elastic and elastoplastic terrain with and without embedded rock.}
\label{fig:plastic_cone_cm}
\end{figure}

The second demonstration example is a dynamic cone penetration test where a cylindrical weight is dropped repeatedly on a cone measuring its penetration depth. 
This is one common way of measuring the mechanical properties of terrain.
The penetration depth of the cone in simulated for two different cases. The first case is a homogeneous material and the second is material with an embedded rock, represented by a rigid body, as illustrated by figure~\ref{fig:plastic_cone}.  The colours 
indicated the magnitude of displacement. The simulated cone penetration is presented in figure \ref{fig:plastic_cone_cm}.
The resistance is higher in the ground with an embedded rock, also before the cone actually come in contact
with the rock.

\section{Conclusions}
\label{sec:conclusions}
A meshfree elastoplastic solid model is made compatible with nonsmooth multidomain dynamics.  
The solid appear as a system of constrained particles in a multibody system on the
same footing as articulated rigid multibodies and power transmission systems.  
The particles carry field variables, e.g., the stress and strain tensor, approximated using the moving least squares method. This method provides a continuous field description throughout the solid.  The dynamic interaction between the deformable solid, rigid
multibodies and other geometric boundaries is modeled by unilateral contact constraints with
dry friction. 
The full system can thus be processed using the same numerical 
integrator and solver framework without introducing additional coupling 
equations with unknown parameters and impulsive behaviour can automatically
be transmitted instantly through the full and strongly coupled system. This enables fast 
and stable simulations of complex mechatronic systems with, or interacting with, elastoplastic
materials.  Demonstration is made with a tracked terrain vehicle driving over 
deformable terrain using the  capped Drucker-Prager model and a cone penetrometer test 
on terrain with and without embedded rock.  

The Jacobian of the deformation constraint is derived and the explicit form is given in Eq.~(\ref{eq:strainjacobiansfinal}).
It can be factored to a constant term that can be pre-computed and multiplied with current particle displacements.
The computational bottleneck of the simulation lies in solving the block sparse mixed linear complementarity problem (\ref{eq:mlcp}).  With 
dedicated hardware, parallel factorisation algorithms or using iterative
solver for a more approximate solution of the terrain dynamics the performance is 
expected to be increased by several order in magnitude.  Exploring this is left
for future work.

The results of numerical experiments of uniform elastic deformations,
presented in figure \ref{fig:elastic_verification} and 
\ref{fig:elastic_hydrostatic_error}, reveal that the solution deviate
from what is expected from analytical solutions.
For example, the confining pressure in 
hydrostatic compression of a cube discretised by $6\times 6\times 6$ particles 
is underestimated by roughly $10$ \% when compressed up to $\Delta V / V =15$ \%.  
The error decrease with reduced time-step and finer spatial discretisation.

The errors are presumably due to the application of \textsc{mls} to the solid dynamics 
equations on strong form.  When the problem involve traction boundary conditions,
meshfree collocation methods suffer 
from poor accuracy and instability \cite{liu:2010:imm}.
The errors are not located solely to the boundaries but influence also the deformations
further in the material, as seen in Figure \ref{fig:elastic_hydrostatic_error}, 
and cause the second principal deviatoric stress invariant to deviate from zero in hydrostatic pressure.

This affect the plastic behaviour also. The permanent plastic deformation
after loading is found to differ by the order of $10$ \% and sometimes the material 
has residual stress that may cause artefacts although none have been observed.
Nevertheless, the developed method is applicable to many type of studies where this
level of accuracy is acceptable and cannot be increased further anyway because of
difficulties in characterising the mechanical properties of the solid material.

This deficiency can be avoided by either using a mixed strong and weak form formulation 
\cite{liu:2010:imm} or applying correction terms of higher order on the
boundary particles \cite{onate:2001:fpm}.  Future work should focus on this issue
to improve the accuracy of the model.  Another area of improvement is to integrate
the plastic yield and flow computations with the mixed complementarity problem for the
constraint forces and velocity update.  This will enable integration with larger 
time-step for strongly coupled problems than the current predictor-corrector method allow.

\section*{Acknowledgements}
The project was funded in part by the Kempe Foundation grant JCK-1109,
Ume{\aa} University and supported by Algoryx Simulation.  The authors are thankful
fruitful discussions and ideas from Prof.~Urban Bergsten at the Swedish University of 
Agriculture, Komatsu Forest and Olofsfors AB.

\section*{Appendix}

\subsection*{A. Capped Drucker-Prager yield surface}
This supplements Section \ref{sec:constraint_elastoplastic} with further details
of the Capped Drucker-Prager yield surfaces in Eq.~(\ref{eq:DP})-(\ref{eq:caps})
illustrated in Figure \ref{fig:gapfunction}.  We follow the smooth cap model
of Dolarevic and Ibrahimbegovic \cite{Dolarevic:2007} with some corrections
to the compression cap.  The three yield functions are
\begin{align}
  \Phi_{\textsc{\tiny DP}}\left( I_1,J_2 \right) &= \sqrt{J_2 } + \frac{\eta}{3} I_1 - \xi c \label{eq:DPapp}\\
  \Phi_{\textsc{\tiny T}}\left( I_1,J_2 \right) &=\left( I_1-T+R_{\textsc{\tiny T}} \right)^2 +J_2 -R^2_{\textsc{\tiny T}} \label{eq:T_cap}\\
  \Phi_{\textsc{\tiny C}}\left( I_1 , J_2,\kappa\right) & =  \frac{\left[ I_1-a\left( \kappa \right) \right]^2}{R_{\textsc{\tiny C}}^{2}}+ J_2 - b\left( \kappa \right)^2 \label{eq:C_cap}
\end{align}
The center, radius and intersection point of the tension cap are 
\begin{align}
  \label{eq:tensioncentre}
C &=\frac{ T-\frac{3\xi c}{\eta}\sin\left( \varphi \right)}{1-\sin\left( \varphi \right)}\\
R_t &= T-C\\
I_t& = \frac{3\xi c}{\eta}-\beta\cos\left( \varphi \right)
\end{align}
where the Drucker-Prager cone angle $\phi = \arctan(\eta)$ in the pressure-shear plane,
 $T$ is the tension cap cutoff and
$\beta = [ \left(3\xi c/\eta - C \right)^2-R_t^2 ]^{1/2}$.  The compression cap
intersection point is
\begin{align}
I_c& = \frac{3\xi c}{\eta}  - \frac{3b\left(\kappa\right)}{\eta\sqrt{1.0 + \left(\frac{\eta}{3}\right)^2R_c^2}}
\end{align}
and $a\left( \kappa \right)$ and $b\left( \kappa \right)$ are the center and main radius of the
compression cap ellipse \cite{Dolarevic:2007}.  The stress gradient of the yield functions are
\begin{align} \label{eq:yield_derivative}
   \frac{\partial \Phi_{\textsc{\tiny DP}}}{\partial \bm{\sigma}} & = 
   \frac{\bm{s}}{2\sqrt{J_2\left( \bm{\sigma} \right)}} + \frac{\eta\left( \phi \right)}{3} \bm{1} \\
   \frac{\partial \Phi_{\textsc{\tiny T}}}{\partial \bm{\sigma}} & = 
   \bm{s} + 2\left( I_1-T+R_t \right)\bm{1}\\
   \frac{\partial \Phi_{\textsc{\tiny C}}}{\partial \bm{\sigma}} & = 
   \bm{s} + \frac{2\left( I_1-a\left( \kappa \right) \right)}{R^2}\bm{1}
\end{align}

\bibliographystyle{wileyj}

\end{document}